\documentclass[acmsmall]{acmart}
\usepackage{mathtools}
\usepackage{subfigure}
\usepackage{graphicx}
\usepackage{amsmath}
\usepackage{multirow}
\usepackage{float}
\usepackage{tikz}
\usepackage{xcolor}
\usepackage{ulem}
\usepackage[linesnumbered,ruled,vlined]{algorithm2e}

\SetCommentSty{algCommentFont}

\AtBeginDocument{%
  \providecommand\BibTeX{{%
    \normalfont B\kern-0.5em{\scshape i\kern-0.25em b}\kern-0.8em\TeX}}}

\SetKwInput{KwInput}{Input}                
\SetKwInput{KwOutput}{Output}              
\SetKwInput{KwEnsure}{Ensure}              
\SetKwRepeat{KwDo}{do}{while}

\setcopyright{none}
\renewcommand\footnotetextcopyrightpermission[1]{}
\acmConference[High Performance Graphics, Poster]{}{July 11--14,
  2022}{}
\acmBooktitle{High Performance Graphics 2022, Poster Session} 

\begin{document}

\title{Adaptive Dynamic Global Illumination}

\author{Sayantan Datta}
\email{sayantan.datta@mail.mcgill.ca}
\orcid{1234-5678-9012}
\affiliation{%
  \institution{McGill Unviversity}
  \state{QC}
  \country{Canada}
}

\author{Negar Goli}
\affiliation{%
  \institution{Huawei/AMD}
  \city{Vancouver}
  \country{Canada}}

\author{Jerry Zhang}
\affiliation{%
  \institution{Huawei}
  \city{Vancouver}
  \country{Canada}
}

\renewcommand{\shortauthors}{Datta et al.}

\begin{abstract}
  We present an adaptive extension of probe based global illumination solution that enhances the response to dynamic changes in the scene while while also enabling an order of magnitude increase in probe count. Our adaptive sampling strategy carefully places samples in regions where we detect time varying changes in radiosity either due to a change in lighting, geometry or both. Even with large number of probes, our technique robustly updates the irradiance and visibility cache to reflect the most up to date changes without stalling the overall algorithm. Our bandwidth aware approach is largely an improvement over the original \textit{Dynamic Diffuse Global Illumination} while also remaining orthogonal to the recent advancements in the technique. \footnote{\href{https://sayan1an.github.io/adgi.html}{Project Page}} \footnote{\href{https://www.highperformancegraphics.org/posters22/HPG2022_Poster5_Adaptive_Dynamic_Global_Illumination.pdf}{Poster}}
\end{abstract}

\ccsdesc[300]{Computing methodologies~Ray tracing}
\ccsdesc[300]{Computing methodologies~Rasterization}

\keywords{Adaptive sampling, irradiance probes, global illumination, real-time}

\begin{teaserfigure}
\vspace{-10pt}
\begin{tikzpicture}
    \node[anchor=south west,inner sep=0] at (0,0){\includegraphics[width=13.5cm, trim={0cm 0.0cm 2.35cm 0cm},clip]{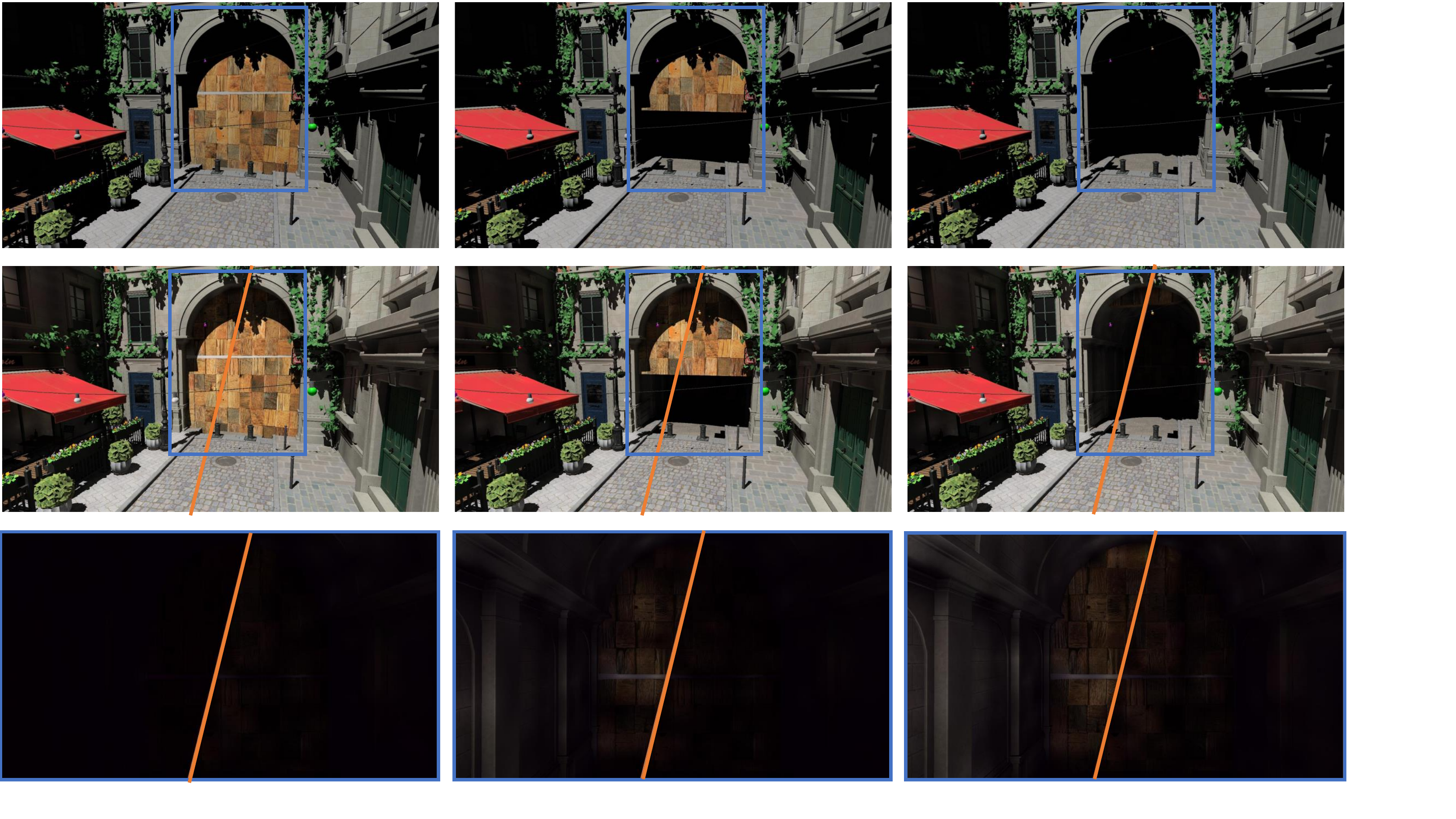}}; 
     \node[anchor=south west] at (1.6,8.1) { \textcolor{black}{\footnotesize $t= \text{start}$} };
     \node[anchor=south west] at (6.0,8.1) { \textcolor{black}{\footnotesize $t= \text{start} + 4s$} };
     \node[anchor=south west] at (10.4,8.1) { \textcolor{black}{\footnotesize $t= \text{start} + 8s$} };
    \node[anchor=south west, rotate=90] at (0.12,6.1) { \textcolor{black}{\footnotesize Direct only} };
    \node[anchor=south west, rotate=90] at (0.10,3.17   ) { \textcolor{black}{\footnotesize Direct + Indirect} };
    \node[anchor=south west, rotate=90] at (0.10,0.15) { \textcolor{black}{\footnotesize Tunnel interior (D + I)} };
    \node[anchor=south west] at (0.05,1.7+3.4) { \textcolor{white}{\footnotesize SSIM/MSE:} };
    \node[anchor=south west] at (0.05,1.7+3.1) { \textcolor{white}{\footnotesize 0.989/0.007} };
    \node[anchor=south west] at (4.6,1.7+3.4) { \textcolor{white}{\footnotesize SSIM/MSE:} };
    \node[anchor=south west] at (4.6,1.7+3.1) { \textcolor{white}{\footnotesize 0.974/0.008} };
    \node[anchor=south west] at (9.15,1.7+3.4) { \textcolor{white}{\footnotesize SSIM/MSE:} };
    \node[anchor=south west] at (9.15,1.7+3.1) { \textcolor{white}{\footnotesize 0.970/0.008} };
    \node[anchor=south west] at (0.06 + 2.0,0.45+2.85) { \textcolor{black}{\footnotesize SSIM/MSE:} };
    \node[anchor=south west] at (0.06 + 2.0,0.45+2.55) { \textcolor{black}{\footnotesize 0.957/0.008} };
    \node[anchor=south west] at (3.85 + 2.8,0.45+2.85) { \textcolor{black}{\footnotesize SSIM/MSE:} };
    \node[anchor=south west] at (3.85 + 2.8,0.45+2.55){
    \textcolor{black}{\footnotesize 0.942/0.009} };
    \node[anchor=south west] at (8.4 + 2.8,0.45+2.85) { \textcolor{black}{\footnotesize SSIM/MSE:} };
    \node[anchor=south west] at (8.4 + 2.8,0.45+2.55) { \textcolor{black}{\footnotesize 0.939/0.010} };
    \node[anchor=south west] at (0.05,2.45) { \textcolor{white}{\footnotesize SSIM/MSE:} };
    \node[anchor=south west] at (0.05,2.15) { \textcolor{white}{\footnotesize 0.993/0.000} };
    \node[anchor=south west] at (4.6,2.45) { \textcolor{white}{\footnotesize SSIM/MSE:} };
    \node[anchor=south west] at (4.6,2.15) { \textcolor{white}{\footnotesize  0.953/0.003} };
    \node[anchor=south west] at (9.15,2.45) { \textcolor{white}{\footnotesize SSIM/MSE:} };
    \node[anchor=south west] at (9.15,2.15) { \textcolor{white}{\footnotesize  0.932/0.004} };
    \node[anchor=south west] at (0.06 + 2.8,0.65) { \textcolor{white}{\footnotesize SSIM/MSE:} };
    \node[anchor=south west] at (0.06 + 2.8,0.35) { \textcolor{white}{\footnotesize 0.837/0.003} };
    \node[anchor=south west] at (3.85 + 3.5,0.65) { \textcolor{white}{\footnotesize SSIM/MSE:} };
    \node[anchor=south west] at (3.85 + 3.5,0.35) { \textcolor{white}{\footnotesize 0.826/0.009} };
    \node[anchor=south west] at (8.4 + 3.5,0.65) { \textcolor{white}{\footnotesize SSIM/MSE:} };
    \node[anchor=south west] at (8.4 + 3.5,0.35) { \textcolor{white}{\footnotesize 0.875/0.010} };
    \node[anchor=south west] at (3.4,-0.2) { \textbf{Left:} Ours@15.6ms\hspace{8mm}\textbf{Right:} Q-DDGI@26.8ms };
\end{tikzpicture}
\vspace{-20pt}
\caption{Our technique demonstrated on a modified \textsc{Bistro Exterior} scene containing $192\times64\times192$ probes. The third row shows the changes inside the tunnel as the gate opens over time. Our techniques responds faster to a dynamic stimuli and offers 1.7-times higher performance compared to the Q-DDGI implementation even with large probe grid containing excess of 2.3 million probes. Q-DDGI, detailed in section \ref{sec:results}, is an extension of vanilla DDGI making it more competitive and comparable against our approach.}
\label{fig:teaser}
\end{teaserfigure}
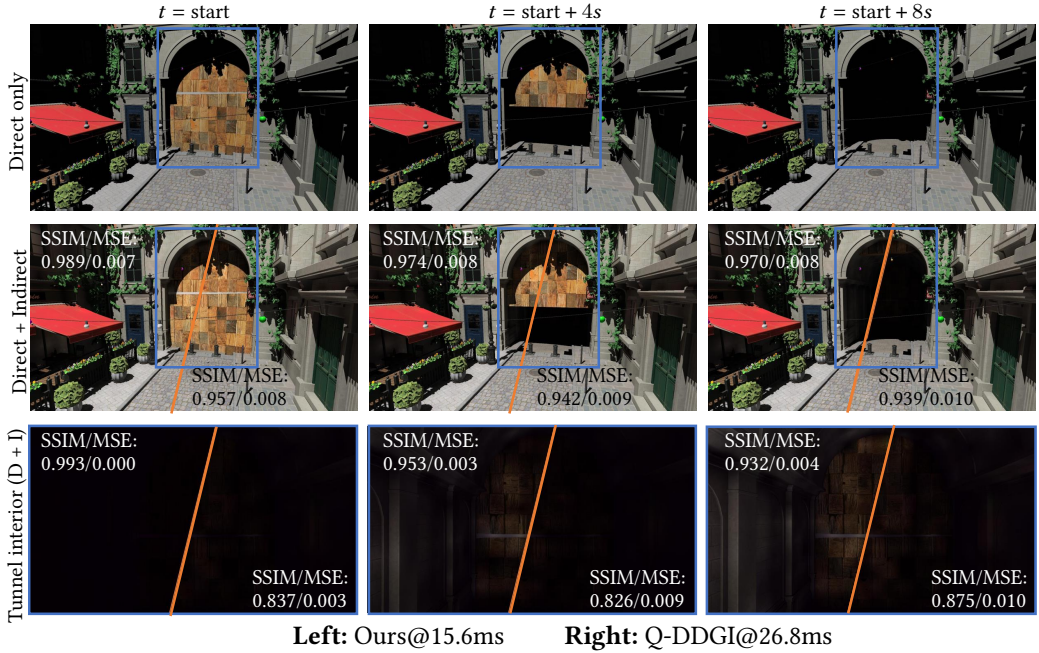

\maketitle

\section{Introduction}

Global illumination (GI) strikingly improves the realism of a virtual scene, but its high computational cost has been a long-standing challenge in its application to real-time rendering \citep{keller2019we}.

Several real-time GI solutions have been proposed, such as screen space \citep{ritschel2009approximating} techniques, which support fully dynamic scenes but suffer from quality issues due to the limited availability of information in screen space. On the other hand, baked texture light-maps only support static geometry but remain popular due to their simplicity, low run-time cost, and quality. Precomputed Radiance Transfer \citep{sloan2002precomputed} combined with light probes \citep{mcguire2017real} and light-maps \citep{iwanicki2017precomputed} solved some of the issues plaguing static light maps; in particular, these approaches support semi-dynamic geometry and self-occlusion while adhering to a strict compute budget. The advent of real-time ray-tracing hardware set the stage for modern fully dynamic GI. Dynamic real-time GI methods build upon the decades of research in sampling, and amortization of shading and visibility across space (pixel/world), angle, and time to improve convergence \citep{RestirGI2021}. Adaptations of several offline techniques such as photon mapping \citep{jensen1996global}, many-light rendering \citep{Lightcut2005, keller1997instant}, and radiosity maps \citep{tabellion2004approximate} have also been explored in the context of modern \citep{VolRestir2021, StoLCuts2020} ray-tracing capable hardware. However, presence of noise in sampled algorithms require the use of strong denoisers. Machine learning denoisers \citep{ChaitanyaDenoise17, NeuralSS} have demonstrable advantages in terms of quality compared to more traditional frequency \citep{AAF12} or variance \citep{SVGF17} based denoisers. However, the prospect of training a neural network, the added complexity of integrating machine learning inference with traditional graphics pipeline, and the proprietary nature of machine learning frameworks have stalled the industry-wide adoption of these techniques. The recent probe-based algorithm, Dynamic Diffuse Global Illumination (DDGI) \citep{DDGI19}, extending the classic irradiance probes, still remains an excellent choice due to its relative simplicity, quality, and cloud streaming capabilities \citep{stengel2021distributed, DDISH-GI}. However, scaling of DDGI in its original formulation is limited, and approaches such as multi-grid hierarchy and probe rolling \citep{majercik2020scaling} are necessary to scale it across large environments. Our adaptive approach focuses on dynamic contents in environments containing millions of probes in a single hierarchy.

We propose Adaptive Dynamic GI (ADGI) algorithm where we trace a few pilot rays per frame to scan the environment and build a coarse representative model of the dynamic events. Using Markov-Chain sampling, we dynamically allocate resources to the critical areas, improving convergence in those regions. While DDGI allocates a fixed number of samples per probe and uniformly distributes samples across directions, ADGI non-uniformly samples the joint spatio-angular domain of the discretized 5D light-field represented by the probes. Our approach essentially decouples resource allocation from the number of probes resulting in a user-controlled performance target (FPS) and improved scaling even with millions of probes. Additionally, our approach results in faster convergence in static and dynamic environments given equal render time. Our approach is drop-in compatible with the original implementation and its several other extensions such as probe rolling and probe volume hierarchies \citep{majercik2020scaling}.
 
We achieve these objectives by formulating a \textit{guided function approximation} technique, which is purposefully accurate in specific regions highlighted by our \textit{guiding function} and thus eliminates the need for uniform resource allocation. Furthermore, we develop a sampling methodology based on temporal Markov-chain, which adapts naturally to a dynamic environment while also enabling scaling across large number of probes. Finally, we discuss memory and bandwidth preserving color compression schemes tailored specifically for our purpose.

\section{Related work}

\textbf{Probe-base approaches:} Modern games rely extensively on light probes for static and dynamic global illumination due to their ease of integration into the game engine pipeline at low run-time cost. Some advocate a uniform grid probe placement due to their simplicity while others have proposed non-unform probes due to their efficiency. Probe based techniques are usually prone to light leakage. As such, uniform grid approaches \citep{mcguire2017real, DDGI19} use additional information, stored in the probes to determine whether a probe is visible from a shade point. Non-uniform approaches may use carefully curated probe placement \citep{wang2019fast} combined with spatial data-structures like octrees to determine the visibility of a probe from a surfel. McGuire et al. \citep{mcguire2017real, DDGI19} stores the depth values of the surrounding geometry from a probe and use a similar idea as Variance-Shadow-Mapping \citep{VSM06} to approximate visibility. However, non-uniform approaches has been mostly limited to static geometry due to their high initial construction cost. Some approaches use rasterization \citep{mcguire2017real, wang2019fast} while other may use ray-tracing \citep{DDGI19} to compute the probe content. Probe based techniques also differ on how they store the information in the probes. Some use discrete textures \citep{mcguire2017real, DDGI19} while other may use a compressed basis representation such as Spherical Harmonics \citep{tatarchuk2005irradiance, DDISH-GI}. Spherical harmonics implicitly pre-filters the content before storage but may cause light and dark ringing issues. Memory bandwidth required for reading and writing from the probes is also a major concern. Texture compression \citep{mcguire2017real, stengel2021distributed} is usually the preferred choice to minimize memory bandwidth. Bandwidth is also crucial for cloud streaming of probe data. In such scenarios, Spherical Harmonics \citep{DDISH-GI} representation may be preferable as they provide excellent compression for low frequency data. At run-time, dynamic probe based \citep{DDGI19} GI solutions uniformly distributes rays across probes to update their content; this quickly becomes a bottleneck as the number of probes increases. Our approach on the other hand, focuses on the optimal distribution of resources to maximize visual fidelity. Various extensions have also been proposed to increase scalablity \citep{majercik2020scaling} of uniform grid approaches such as multiple-volume hierarchies and probe rolling. Our approach remains largely orthogonal and fully compatible with these extensions.

\textbf{Adaptive sampling:} Adaptive sampling has been used in the context of screen-space ray-traced global illumination where more samples are accumulated in regions with high noise and high frequency \citep{toshiyaAdaptive08}. Adaptive sampling is also useful for filtering soft shadows \citep{AAF12}, where pilot-rays model the spatial frequency of shadow-penumbra and provide the number of additional samples required at each pixel to improved convergence. Neural versions \citep{neuralAdaptive20} of adaptive sampling has also been proposed where a neural network generates a sampling-map that is tightly coupled to a post-process neural-denoiser. Conceptually our approach is similar, but our execution is tailored for the problem of temporally coherent sampling of probes. We refer readers to section \ref{sect:relatedWorkExtension} for an extend related work in irradiance-caching, screen-space GI and MCMC techniques

\section{Overview \label{sec:review}}

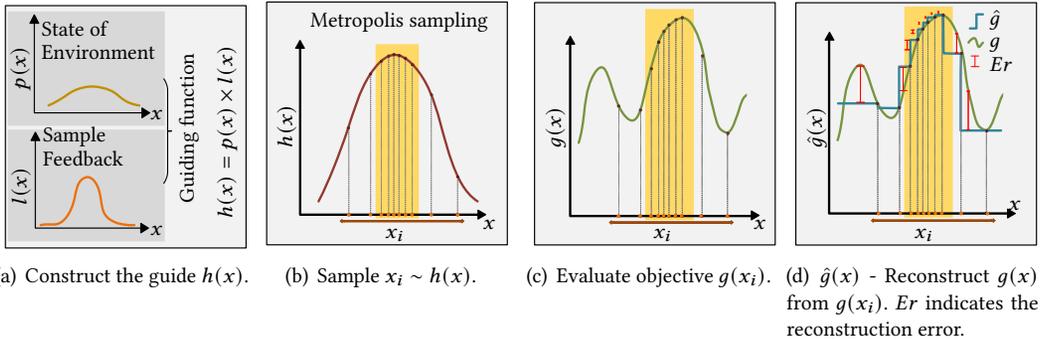
\begin{figure*}
\begin{center}
\centering
\subfigure[ Construct the guide $h(x)$.\label{fig:graph_h}]{ \begin{tikzpicture}
    \node[anchor=south west,inner sep=0] at (0,0) {
    \includegraphics[scale=1.75]{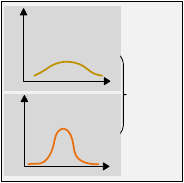}};
    \node[anchor=south west, rotate=90] at (2.74, 0.51) { \textcolor{black}{\footnotesize Guiding function} };
    \node[anchor=south west, rotate=90] at (3.25, 0.31) { \textcolor{black}{\footnotesize $h(x)= p(x) \times l(x)$} };
    \node[anchor=south west, rotate=90] at (0.5, 2.0) { \textcolor{black}{\footnotesize $p(x)$} };
     \node[anchor=south west, rotate=90] at (0.5, 0.48) { \textcolor{black}{\footnotesize $l(x)$} };
 \node[anchor=south west] at (1.85, 01.6) { \textcolor{black}{\footnotesize $x$} };
     \node[anchor=south west] at (1.85, 0.08) { \textcolor{black}{\footnotesize $x$} };
     
     \node[anchor=south west] at (0.4, 2.7) { \textcolor{black}{\footnotesize State of} };
     \node[anchor=south west] at (0.4, 2.4) { \textcolor{black}{\footnotesize Environment} };
      \node[anchor=south west] at (0.42, 1.25) { \textcolor{black}{\footnotesize Sample} };
    \node[anchor=south west] at (0.42, 1.0) { \textcolor{black}{\footnotesize Feedback} };
\end{tikzpicture}}\hspace{1.pt}
  \subfigure[Sample $x_i \sim h(x)$. \label{fig:samp_h}]{ \begin{tikzpicture}
    \node[anchor=south west,inner sep=0] at (0,-0.2) {
    \includegraphics[scale=1.75]{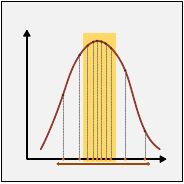}};
    \node[anchor=south west] at (2.8, 0.12 - 0.2) { \textcolor{black}{\footnotesize $x$}};
    \node[anchor=south west] at (0.5, 2.72 - 0.2) { \textcolor{black}{\footnotesize Metropolis sampling}};
     \node[anchor=south west] at (1.5, -0.04 - 0.2) { \textcolor{black}{\footnotesize $x_i$}};
    \node[anchor=south west, rotate=90] at (0.56, 1.2 - 0.2) { \textcolor{black}{\footnotesize $h(x)$} };
\end{tikzpicture}} \hspace{1.pt}
  \subfigure[Evaluate objective $g(x_i)$. \label{fig:graph_g}]{ \begin{tikzpicture}
    \node[anchor=south west,inner sep=0] at (0,0) {
    \includegraphics[scale=1.75]{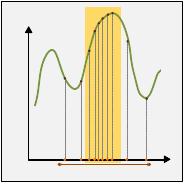}};
    \node[anchor=south west] at (2.84, 0.12) { \textcolor{black}{\footnotesize $x$} };
    \node[anchor=south west] at (1.5, -0.04) { \textcolor{black}{\footnotesize $x_i$}};
    \node[anchor=south west, rotate=90] at (0.58, 1.2) { \textcolor{black}{\footnotesize $g(x)$} };
\end{tikzpicture}}\hspace{1.pt}
  \subfigure[$\hat{g}(x)$ - Reconstruct $g(x)$ from $g(x_i)$. $Er$ indicates the reconstruction error. \label{fig:graph_g_hat}  ]{ \begin{tikzpicture}
    \node[anchor=south west,inner sep=0] at (0,0) {
    \includegraphics[scale=1.75]{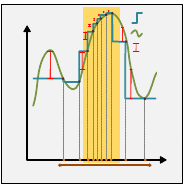}};
   \node[anchor=south west] at (2.75, 0.38) { \textcolor{black}{\footnotesize $x$} };
    \node[anchor=south west] at (1.5, -0.04) { \textcolor{black}{\footnotesize $x_i$}};
    \node[anchor=south west] at (2.5, 2.2) { \textcolor{black}{\footnotesize $Er$} };
    \node[anchor=south west] at (2.5, 2.45) { \textcolor{black}{\footnotesize $g$} };
    \node[anchor=south west] at (2.5, 2.75) { \textcolor{black}{\footnotesize $\hat{g}$} };
    \node[anchor=south west, rotate=90] at (0.578, 1.2) { \textcolor{black}{\footnotesize $\hat{g}(x)$} };
\end{tikzpicture}}
 \end{center}
 \vspace{-15pt}
  \caption{\label{fig:graph}Figure showing the steps in our adaptive-sampling strategy. We define a guiding function $h(x)$ that highlights (in yellow) the interesting regions of the domain. The samples $x_i$ obtained from $h(x)$ are used to evaluate the objective $g(x_i)$. Our goal is to obtain an approximate representation of $g(x)$, denoted as $\hat{g}(x)$, from the $(x_i, g(x_i))$ pairs. As more samples are obtained from the highlighted region, the reconstruction error is lower in the yellow area, as shown in sub-figure (d).}
\end{figure*}

We focus on two primary issues with DDGI in its original formulation. First, the technique does not allow for the non-uniform allocation of resources, resulting in unnecessary probe updates in regions that are not crucial for visual fidelity. Seconds, it does not update the probes quick enough to reflect transient changes in the scene environment. Our adaptive strategy involves detecting the changes in the environment and allocating resources driven by the detected changes. While the detection phase requires allocating additional resources, our empirical evaluations suggest our non-uniform adaptive sampling compensates for the lost efficiency in the detection phase. Our detection phase also enables fast probe updates for capturing transient changes in the scene. We model our technique as \textit{guided function approximation} where we approximate a continuous function (e.g. 5D light-field) using a discrete (e.g. probes) representation driven by a \textit{guiding function}.  

A naive approach to approximate a continuous function is to discretize the domain and reserve a representative sample for each discretization. The strategy is useful when the domain is relatively small; however, as the domain gets larger or the number of discretizations increases, it is prohibitively expensive to update all discretizations in real-time. This is one of the issues plaguing the original DDGI technique. In many applications, it is not necessary to update the entire domain uniformly; instead, we can tolerate more approximation errors in some regions than others. A simple example is foveated rendering, where errors in the periphery are less intrusive than those near the gaze center. In our case, we need the most accuracy in probes contributing to final shading.

We introduce the notion of \textit{guiding function}, which highlights the regions where a higher reconstruction accuracy is desired. We define the guide using a product of terms - the first term represents the current state of the environment while the second term is a feedback from the sampled cache. We sample the guide using a temporally coherent Markov-chain and use the samples to update our approximate representation using a parallel thread-safe approach. Thus our approach is summarized in three steps - defining a guiding function, sampling the guide, and using the samples to update the approximate representation. We describe these steps in sections \ref{subsec:GF}, \ref{subsec:SMC} and \ref{subsec:approx} while we discuss various implementation specific details in section \ref{sec:implementationDetails}. See figure \ref{fig:graph}.

Our approach provides two distinct advantages compared to the original DDGI - approximation quality and scalability. At any time, we concentrate our resources on a potentially challenging area as opposed to the entire domain. Provided our guide correctly identifies the challenging regions, the quality is improved due to a higher concentration of resources in the appropriate region. Since we sample the guide independent of the number of discretizations, the decoupling allows for a high number of statically allocated probes without affecting run-time performance. Increased discretizations improve approximation quality while the independence of sampling from the number of discretizations improves scalability. More specifically, we transparently increase the number of discrete probes without affecting performance. The run-time performance depends on the number of samples we generate; the samples are channeled to the appropriate areas by the guiding function. Our Markov-chain sampling is highly parallel, temporally coherent, and scalable, making it suitable for real-time temporally distributed reconstruction of large probe grids.

\subsection{Background}
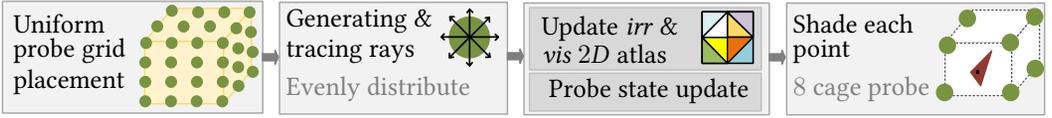
\begin{figure*}
\begin{tikzpicture}
    \node[anchor=south west,inner sep=0] at (0,0) {
   \includegraphics[scale=2.10]{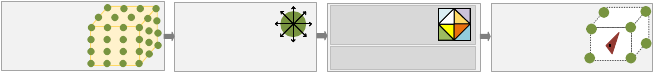}};
       \node[anchor=south west] at (0.07, 1.0) { \textcolor{black}{Uniform } };
       \node[anchor=south west] at (0.07, 0.60) { \textcolor{black}{ probe grid} };
      \node[anchor=south west] at (0.07, 0.20) { \textcolor{black}{placement} };

            \node[anchor=south west] at (3.7, 1.0) { \textcolor{black}{Generating \&} };
             \node[anchor=south west] at (3.7, 0.6) { \textcolor{black}{tracing rays} };
             \node[anchor=south west] at (3.7, 0.1) { \textcolor{gray}{Evenly distribute}};
           
 
             \node[anchor=south west] at (7.2, 0.08) { \textcolor{black}{Probe state update} };
             
                           \node[anchor=south west] at (7.1, 0.9) { \textcolor{black}{Update \textit{irr} \&} };
                              \node[anchor=south west] at (7.1, 0.6) { \textcolor{black}{\textit{vis} $2D$ atlas} };

                   \node[anchor=south west] at (10.45, 1.0) { \textcolor{black}{Shade each}};
                   
                   \node[anchor=south west] at (10.45, 0.60) { \textcolor{black}{
                   point}};
                   
                   \node[anchor=south west] at (10.45, 0.1) { \textcolor{gray}{
                     $8$ cage probe}
                     };
 
               
                  
\end{tikzpicture}
\caption{The figure illustrates the main steps of DDGI algorithm. Algorithm defines a uniform grid of probes and trace uniform-random rays in all direction from each probe. Based on the hit information, we compute the visibility (vis) and irradiance (irr) and update the $2D$ atlas. We also update the probe states based on visibility information (back-face hit ratio). Finally, for each shade-point, we query the eight bounding probes surrounding it and interpolate them to compute incoming indirect illumination.}
\label{fig:DDGI}
\end{figure*}

Here we briefly describe the original DDGI algorithm. DDGI consists of a 3D grid of directionally resolved irradiance probes that are updated in real-time through hardware ray-tracing. The probes also contains visibility information to prevent light leakage. The probe representation has many benefits, it performs optimally for diffuse indirect transport and is relatively inexpensive to encode and decode information to and from the probes. The algorithm evenly distributes ray-samples outwards from the probe center at each active probe in a stochastic rotated spiral pattern. DDGI is a two step algorithm. First, it updates the shading on the probe texels. Next a screen-space pass where the up-to-date probe content is used for shading the camera-pixels. The probe texel values are encoded into a spherical-mapped diffuse irradiance-texture with $8 \times 8$ resolution. Probes also captures the average ray-hit distance, and squared distances to the nearest geometry at $16 \times 16$ resolution. DDGI temporally filters the probe texels by blending in the new values using a fixed hysteresis. The visibility data is used to decide whether a probe is visible at a shade-point and also used to infer whether a probe is inside a geometry and deactivated. The probe's state is not limited to on or off and can vary with scenarios \citep{majercik2020scaling}. The world-position of the screen-space pixel is used as a key to the probe-texture lookup. The lookup interpolates the corresponding eight probes of the grid voxel containing the shade-point. The algorithm is illustrated in figure \ref{fig:DDGI}. DDGI algorithm is suitable for diffuse and slow changing phenomena in time. Therefore DDGI, combined with our adaptive-sampling strategy is a reasonable real-time GI approximation for dynamic scenes.

\subsection{Guiding function \label{subsec:GF}} 
  
As summarised in section \ref{sec:review} and figure \ref{fig:graph}, a \textit{guiding function} highlights the important areas in the domain, i.e., challenging regions where more resources are required. These highlighted areas receive more adaptive samples, reducing the approximation error in those regions. Mathematically, the domain of the guiding function $h:R^d \rightarrow R$ is the continuous 5D light field. Upon query, the guide function returns a scalar value indicating the importance of a sampled point. In our case, $d = 5$ as the domain is a 5-dimensional space of world-space positions and directions, and the guide encodes the importance of sampling a direction on a probe (texel's importance).

We model the guiding function ($h$) as a product of two terms. The first term, we call $f:R^d \rightarrow R$, represents the value in sampling a texel based on our understanding (limited) of whether such a texel would contribute towards the final screen-space shading. The second term is the observed sampled evidence (a.k.a irradiance cache) as they become available. Initially, the irradiance cache is empty but filled progressively through sampling. We define the first term based on some heuristics that describes our understanding of the probe-environment:

\begin{itemize}
\item Probes closer to the camera,
\item Probes closer to geometric surfaces,
\item Directions on the probes facing away from geometric surfaces,
\item Directions on the probe with higher incoming irradiance,
\item Directions with temporal change in irradiance and visibility 
\end{itemize}

We trace \textit{pilot rays} from the probes to generate the information necessary to quantify the above heuristics. We also call it the \textit{detection phase} where we pre-scan the scene environment for changes. We denote the individual heuristics as $f_i:R^d \rightarrow R$, and compose them into its final form $f$ as shown in equation \ref{eq:guide}, where $\phi$ represents a composition function. The composition function is simply a recipe to appropriately combine the individual heuristics. We quantify the individual heuristics ($f_i$) in section \ref{sec:priorConstr} and the composition ($\phi$) in section \ref{sec:priorCompose}.

\begin{equation}
\label{eq:guide}
 \begin{aligned}     
 f = \phi(f_0, f_1,..., f_i).
 \end{aligned}
\end{equation}


The second term uses the stored irradiance in the probes, denoted by $\hat{g}$, to modulate the first term. We model the second term as -
 $exp\left(\alpha \cdot \hat{g}(x)/f(x)\right)$, where the scalar $\alpha \in [0, \infty)$ indicates our confidence in the irradiance probe content; a higher value indicating greater confidence. Note that a stored texel with high irradiance value may or may not have a high contribution to the final shading. Example - in a dynamic environment the probe content from the last frame is quickly outdated and thus less useful. The parameter $\alpha$ models this uncertainty. \textcolor{black}{The term $f(x)$ in the denominator ensures that we only trust $\hat{g}(x)$ when $f(x)$ is low.} Finally, we define the guiding function as:

 \begin{equation}
 \label{eq:DDGIBays}
 h(x) = {exp\left(\alpha \cdot \frac{\hat{g}(x)}{f(x)}\right)} \cdot {f(x)}.
\end{equation}
 
\subsection{Sampling the guide\label{subsec:SMC}}  

Next we sample the guiding function (equation \ref{eq:DDGIBays}). Mathematically, given an unnormalized distribution $h:R^d\rightarrow R$, our goal is to obtain samples $x_i$ from $h(x)$, where $x_i \in R^d$.

Our sampling algorithm is straightforward. We use the Metropolis sampling, as shown in algorithm \ref{alg:MarkovChainMT} to sample $h$. The algorithm randomly initializes a state ($x_0 \in R^d$) and moves the state forward based on the acceptance of a newly proposed state. We generate the proposed states by perturbing the current state with a zero-mean Gaussian noise, also known as \textit{Random-walk} [\ref{alg:MetropolisStep}]. 

\begin{algorithm}[t!]
\DontPrintSemicolon
\caption{Metropolis algorithm}\label{alg:MarkovChainMT}
\KwInput{$h$: Guide distribution, $M$ : No. of iterations}
\KwInput{$K$: No. of initial samples to reject}
\KwOutput{$x$ : Sample}
\KwEnsure{$M \geq 2$, and $K < M$}
$j \gets$ ShaderInvocationIndex()\;
$x_0 \gets S[j]$ \tcp*{Initialize Markov-chain from memory}
\While{$i \gets 0$ to $M-1$}
{
    $x_{i+1} \gets$ RandomWalk$ \: (x_i, h(x_i))$ \tcp*{Random walk step, algorithm \ref{alg:MetropolisStep}}
    \If{$i > K$}
    {
        \tcc{Use sample $x_{i+1}$ for probe updates, see algorithm \ref{alg:ApproxAlgo}}
    }
}
$S[j] \gets x_i+1$ \tcp*{Save Markov-chain state}
\end{algorithm}


\textbf{Parallelism:} Note that algorithm \ref{alg:MarkovChainMT} runs as a shader invocation, meaning several instances of the chain run in parallel. Each instance is independent with its own memory to load and store the chain state (denoted by S[] in algorithm \ref{alg:MarkovChainMT}). The instances generate thousands of samples per frame. As an input to our algorithm, we explicitly specify the number of chains that run in parallel, thus controlling the number of adaptive samples and performance. Contrasting with the original DDGI, the number of samples in the original implementation is proportional to the number of probes which increases cubically with scene dimensions. As such, it is difficult to scale up when the scene gets larger or when using a denser probe grid. Our approach is independent of the discretization resolution and scales better to higher probe counts without compromising approximation quality. 

\textbf{Mixing-time:} Initially, a Markov chain requires many iterations for the chain to generate samples from the target distribution (here $h(x)$), a phenomenon known as mixing time. We avoid this problem by bootstrapping the initial chain state from the last frame. As such, we keep the number of iterations per frame small, but over frames, the chain effectively accrues many iterations. 

\textbf{Distribution stationarity:} Markov chain sampling requires the target distribution $h(x)$ remain stationary. Due to a dynamic scene environment, the stationarity condition is seemingly violated. This may affect the approximation quality of our technique if the distribution changes rapidly between frames. However, we have several contingencies to deal with the issue. First, we target high frame-rates, which minimizes the change in the target distribution between consecutive frames. As an additional margin of safety, we reject initial $K$ samples per frame as shown in algorithm \ref{alg:MarkovChainMT}, line 5. This ensures our usable samples are obtained closer to the target distribution. Note that the evaluation time for $h(x)$ negligible and thus rejecting few initial samples per frame does not significantly impact performance. We also smooth out the target distribution (see section \ref{subsect:visChange}) using spatio-temporal convolution to minimize abrupt changes in the target across frames. 

\textbf{Temporal tracking:} Since our target distribution may vary with time, we require the samples generated from the Markov-chain to closely follow the distribution to capture the transient changes in the environment. We make some crucial modifications to our sampling algorithm to allow for fast tracking of the target distribution, which we discuss in detail in section \ref{subsect:samplingDetails}.

\begin{table*}[t!]
\caption{\label{tab:listofsyms}List of symbols}
\vspace{-10pt}
\begin{tabular}{|c|c|c|}
\hline
Symbol & Description & Remarks \\ \hline
$f$ & Heuristics model & Section \ref{subsec:GF} \\ \hline
$h$ & Guiding function/Target distribution & Section \ref{subsec:GF}, \ref{subsec:SMC} \\ \hline
$g$ & Objective function & \begin{tabular}[c]{@{}c@{}}Symbolic proxy for $g_r$, $g_c$.\\ Section \ref{subsec:approx} \end{tabular} \\ \hline
$\hat{g}$ & Approximation of objective function & \begin{tabular}[c]{@{}c@{}}Symbolic proxy for $\hat{g_r}$, $\hat{g_c}.$\\ Section \ref{subsec:approx}\end{tabular} \\ \hline
$g_r$ & 5D Light field & Section \ref{subsec:approx} \\ \hline
$g_c$ & Chebychev visibility & Section\ref{subsec:approx} \\ \hline
$\hat{g_r}$ & \begin{tabular}[c]{@{}c@{}}Approximation of 5D light field\\ (Irradiance cache)\end{tabular} & Section \ref{subsec:approx}, \ref{subsect:probeIrCach} \\ \hline
$\hat{g_c}$ & \begin{tabular}[c]{@{}c@{}}Approximation of Chebychev visibility\\ (Visibility cache)\end{tabular} & Section \ref{subsec:approx}, \ref{subsect:probeVisCach}  \\ \hline
$x$ or $x_i$ & Markov-chain samples & \begin{tabular}[c]{@{}c@{}}Symbolic proxy for $p_i$, $\omega_i$.\\ Section \ref{subsec:SMC} \end{tabular} \\ \hline
$p_i$ & Positional ($\in R^3$) component of $x_i$ & -- \\ \hline
$\omega_i$ & Directional ($\in R^2$) component of $x_i$ & -- \\ \hline
\end{tabular}
\end{table*}

\subsection{Approximation \label{subsec:approx}}
With samples obtained from the highlighted (figure \ref{fig:samp_h}) parts of the domain, we focus on using the samples to evaluate (figure \ref{fig:graph_g}) and reconstruct (figure \ref{fig:graph_g_hat}) our objective function. The term \textit{objective function} refers to the quantity we aim to approximate. Mathematically, we denote our objective function as $g:R^d \rightarrow R^c$, and its approximate reconstruction as $\hat{g}$.  For ADGI, we have two objective functions - the light field $g_r:R^5 \rightarrow R^3$, and Chebychev-visibility $g_c:R^5 \rightarrow R^2$ surrounding the probes. We denote their approximate reconstructions as the irradiance cache $\hat{g_r}$, and the visibility cache - $\hat{g_c}$ respectively. See section \ref{subsect:probeIrCach} and \ref{subsect:probeVisCach} for more details.

\textbf{Updating $\mathbf{\hat{g}}$:} We evaluate the continuous objective function $g$ at collected sample points $x_i$ and store the evaluations - $g(x_i)$ into $\hat{g}$,  as shown in algorithm \ref{alg:ApproxAlgo}. For ADGI, the evaluation step involves tracing a ray to query the local light field and visibility. \textcolor{black}{At each Metropolis iteration, the evaluated samples update the closest entry in the probes ($\hat{g}$) within a critical section construct.} 
\begin{algorithm}[t!]
\caption{Approximation algorithm}\label{alg:ApproxAlgo}
\DontPrintSemicolon
\KwInput{$x$: Markov-chain samples}
\SetKwFunction{updRpr}{UpdateRepresentation}
\SetKwProg{Fn}{function}{:}{}
  \Fn{\updRpr{$x$}}
  {
    $v \gets g(x)$ \tcp*{Evaluate sample, ray-trace}
    AtomicMovingAvg$(x, v)$ \tcp*{Populate $\hat{g}$, see algorithm \ref{alg:atmMvAvgUpd}}
  }
\end{algorithm}

\textbf{Representing $\mathbf{\hat{g}}$:} Prior work represent  $\hat{g}$ as either as discrete LUTs \citep{DDGI19}, continuous Spherical Harmonics \citep{DDISH-GI}, Neural Networks \citep{NRC2020}, or any combination. In our case, the choice to use a discrete representation is based on several factors. First, multiple parallel streams of Markov-chain samples may update the same memory location in $\hat{g}$. As such, provisions are necessary to prevent race conditions. We also need a representation that handles temporal accumulation and quickly update itself to reflect any transient changes in the scene. Finally, the representation must be bandwidth efficient to improve the read and write performance. We refer to section \ref{subsect:probeIrCach} and \ref{subsect:samplingDetails} for details.

\subsection{\textcolor{black}{MCMC analysis} \label{subsec:mcmcAnalysis}}

In this section, we analyze our adaptive sampling algorithm in the context of MCMC (Markov Chain Monte Carlo). Note that our goal is not variance reduction through importance sampling; rather the focus is guided approximation of the objective function via sampling the target function. As such, unlike importance sampling, the sampling function is not necessarily correlated to the integrand. With this distinction in mind, we first look at the equation driving importance sampling using MCMC and then repurpose it for guided function approximation.

The following equation shows a typical case of importance sampling where the objective is to compute the integral $\int h(x)g(x)dx$ and there exists a strategy to sample from h(x). In many typical scenarios (e.g. full Bayesian inference), the distribution $h(x)$ is a proper distribution ($\int h(x)dx=1$) but does not have an efficient sampling mechanism. This where Markov Chain MC is useful.        

\begin{equation}
\int h(x)g(x)dx \approx \left\{ \frac{1}{M} \sum\limits_{i=0}^{M-1} g(x_i) \right\} \int h(x)dx,\;\; x_i \sim h(x).
\label{eq:MCMCsampling}
\end{equation}

In contrast, our choice of Markov Chain (Metropolis) is primarily technical - simplicity, GPU parallelism and temporal sample tracking. Nevertheless, the same equations provide meaningful insight - albeit in a different context of adaptive sampling. In our algorithm, we simply sum the samples obtained from the target distribution without taking into account the sample density. This is equivalent to computing the following:

\begin{equation}
    I = \frac{1}{M} \sum\limits_{i=0}^{M-1} g(x_i),\;\; x_i \sim h(x).
\end{equation}

While our goal is to estimate $\int_\Omega g(x)dx$, the expectation of $I$ (rearranging equation \ref{eq:MCMCsampling}) is:

\begin{equation}
    \mathbb{E}\left[ I \right] = \frac{\int_\Omega h(x)g(x)dx}{\int_\Omega h(x) dx},
\end{equation}

where $\Omega$ is the domain of integration. Clearly, the expected value of $I$ does not converge to the correct estimate - $\int_\Omega g(x)dx$. However, there are two factors to consider - size of the domain $\Omega$ and shape of $h(x)$ in the domain. First consider the limit case where $\Omega \to 0$. In this case, the integrals collapses to a point evaluation and indeed the expected value of $I$ equals the unbiased estimate as shown below.
\begin{equation}
    L.H.S. = \lim_{\Omega \to 0}\frac{\int_\Omega h(x)g(x)dx}{\int_\Omega h(x)dx} =\frac{\int_\Omega h(x)g(x)\delta(x-x_0)dx}{\int_\Omega h(x)\delta(x-x_0)dx} = g(x_0).
    \label{pointDomain1}
\end{equation}
\begin{equation}
    R.H.S. = \lim_{\Omega \to 0} \int_\Omega g(x)dx  = \int_\Omega g(x)\delta(x-x_0)dx = g(x_0).
\end{equation}

In the above equation, $\delta$ is the Kronecker delta. The result is important as it shows with increasing probe resolution, bias is reduced. However, reducing texel size is not always practical as more rays and memory are required to populate and store a high resolution probe. Notice how the term $h(x)$ is cancelled in equation \ref{pointDomain1}. When the domain of integration is sufficiently small, $h(x)$ is practically constant and the term cancels out in the denominator and numerator.      

We now consider the shape of $h(x)$. While the target h(x) varies globally, it is piece-wise constant at a local scale due to its tabular nature. More crucially, the target $h(x)$ is stored at a much lower resolution compared to the irradiance probe $\hat{g}(x)$. This implies $h(x)$ is practically constant across a texel of the irradiance probe. The expected value of $I$ for the $k^{th}$ texel is thus given by:
\begin{equation}
    \mathbb{E}\left[ I_k \right] = \frac{\int_{T_k} h(x)g(x)dx}{\int_{T_k} h(x) dx} = \frac{\int_{T_k} c_kg(x)dx}{\int_{T_k}c_k dx} = \frac{\int_{T_k} g(x)dx}{\int_{T_k}dx},
\end{equation}

where $T_k$ represents the domain of $k^{th}$ texel and $c_k$ represents the piece-wise constant value of $h(x)$ when $x \in T_k$. The area estimate $\int_{T_k} dx$ is fixed for all texels and equivalent to $4\pi/\#resolution$. Thus, due to the tabular nature of our target function, the estimates of irradiance texels remain un-biased. While performing texture filtering over irradiance texels, it is possible to compute an unbiased estimate by weighing the texel values with $c_k$ as follows:
\begin{equation}
    I_k^{filter} = \sum\limits_{j \in \mathcal{N}_k} w_{k-j}I_{k-j}, \;\; w_i = c_i/\sum\limits_{j \in \mathcal{N}_k}c_j,
\end{equation}

where $\mathcal{N}_k$ represents the texels in the neighbourhood of texel $k$. The values $c_i$ are obtained by querying the probes storing $h(x)$. Note that bias is unavoidable as we blend samples temporally in a dynamic environment. In a dynamic environment, the objective is evolving and the bias manifests itself as temporal lag. Practically however, within a small time window, both $h(t)$ and $g(t)$ are assumed constant and the samples can be blended using a windowed moving average. Note that windowed moving average requires storing historical information. A cheaper but biased approximation to windowed moving average is exponential moving.

\begin{figure*}
\begin{tikzpicture}
    \node[anchor=south west,inner sep=0] at (0,0) {
   \includegraphics[scale=2.2]{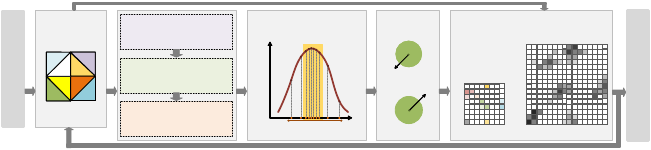}};
   \node[anchor=south west, rotate=90] at (0.6, 0.95) { \textcolor{black}{\footnotesize World space} };
    \node[anchor=south west] at (0.9, 2.6) { \textcolor{black}{\footnotesize Trace $8$} };
      \node[anchor=south west] at (0.9, 2.2) { \textcolor{black}{\footnotesize pilot rays} };
       \node[anchor=south west] at (0.8, 0.6) { \textcolor{black}{\footnotesize Octahedral} };
      
       \node[anchor=south west] at (2.72, 2.6) { \textcolor{black}{\footnotesize Quantify heuristics} };
       \node[anchor=south west] at (2.95, 2.23) { \textcolor{gray}{\footnotesize $\phi(f_0, f_1,..., f_i) $} };
            \node[anchor=south west] at (2.85, 1.65) { \textcolor{black}{\footnotesize Model feedback} };
             \node[anchor=south west] at (3.05, 1.25) { \textcolor{gray}{\footnotesize $exp(\alpha \cdot \hat{g} / f)$} };
              \node[anchor=south west] at (2.9, 0.7) { \textcolor{black}{\footnotesize Guide function} };
             \node[anchor=south west] at (3.0, 0.27) { \textcolor{gray}{\footnotesize $p(x) \times l(x)$} };
              \node[anchor=south west] at (5.51, 2.6) { \textcolor{black}{\footnotesize Metropolis sampling}
               };
             \node[anchor=south west, rotate=90] at (6.05, 1.0) { \textcolor{black}{\footnotesize $h(p,\omega)$} };
              \node[anchor=south west] at (6.45, 0.16) { \textcolor{black}{\footnotesize $(p_i,\omega_i)$}};
             \node[anchor=south west] at (8.4, 0.15) { \textcolor{black}{\footnotesize Irradiance} };
             \node[anchor=south west] at (8.45, 1.28) { \textcolor{black}{\footnotesize Visibility} };
             
              \node[anchor=south west, rotate=45] at (9.5, 0.7) { \textcolor{black}{\footnotesize $\omega_i$} };
               \node[anchor=south west, rotate=45] at (8.6, 1.6) { \textcolor{black}{\footnotesize $-\omega_i$} };
                 \node[anchor=south west] at (9.1, 1.95) { \textcolor{black}{\footnotesize $p_i$} };
                   \node[anchor=south west] at (8.89, 0.48) { \textcolor{black}{\footnotesize $p_i$} };
               
                   \node[anchor=south west] at (8.4, 2.55) { \textcolor{black}{\footnotesize Trace ray}
               };
               
                   \node[anchor=south west] at (10.06, 2.60) { \textcolor{black}{\footnotesize Update irradiance, visibility }
               };
                \node[anchor=south west] at (10.06, 2.37) { \textcolor{black}{\footnotesize cache}
               };
                     \node[anchor=south west] at (9.95, 0.09) { \textcolor{black}{\footnotesize Irradiance ($\hat{g_r}$)} };
                     \node[anchor=south west] at (10.38, 1.41) { \textcolor{black}{\footnotesize $8\times8$} };
                      \node[anchor=south west] at (11.9, 0.08) { \textcolor{black}{\footnotesize Visibility ($\hat{g_c}$)} };
                       \node[anchor=south west] at (12.24, 2.32) { \textcolor{black}{\footnotesize $16\times16$} };
               
               
               \node[anchor=south west, rotate=90] at (14.45, 0.50) { \textcolor{black}{\footnotesize Deferred shading}
               };
              
\end{tikzpicture}
\caption{This figure illustrates our overall algorithm. We trace 8 pilot rays, one from each octant on the probe and approximate the heuristic model $f(p, \omega)$. Using the heuristic and feedback, we define the guide $h(p, \omega)$ and sample it using Metropolis sampling. The sampled ($p_i, \omega_i$) are used to trace more adaptive ray samples, gathering hit-distance and irradiance at the sample points. We update the probe-cache ($\hat{g}$) with adaptive-samples. The cache is used in the next shader and also looped back as feedback to model the target.}
\label{fig:Method}
\end{figure*}
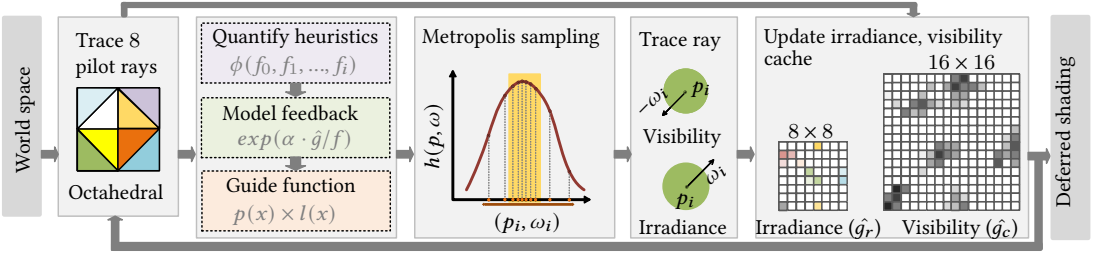 

\section{Implementation details \label{sec:implementationDetails}} This section provides the several implementation details with a brief summary in figure \ref{fig:Method}.

\subsection{Heuristics construction \label{sec:priorConstr}} The section describes the construction of $f$ using the heuristics discussed in section \ref{subsec:GF}. Our goal is to measure and quantify the heuristics that highlight the probes which actively contribute to the final shading and require additional resources for faster convergence. We represent the heuristics either parametrically (equation \ref{eq:cameraD}) or using an explicit LUT representation as shown in figure \ref{fig:probeStructure}(a). The LUT is constructed such that each probe has eight texels corresponding to an octant. We trace a ray for each octant; the rays return the hit distance and incoming irradiance at the hit-point. From this information, we compute several quantities (equation \ref{eq:ProbeVis} - \ref{eq:irradChange}) and store them in the LUT/texture mapped to the probe octants. We define and evaluate the following heuristics for a probe at position $p$ and a direction $\omega$.

\subsubsection{Distance from camera} A probe far away from the camera is less likely to contribute to the final shading. We represent this parametrically as described in equation \ref{eq:cameraD}, where $p$ represents probe position, $c$ camera position and $k$ is a threshold set by the user. 
\begin{equation}
  {f_c(p, \omega)} =
  \begin{cases}
    1 & \text{if $||p - c|| < k$ }, \\
    e^{-(||p - c|| - k)} & \text{otherwise}.
  \end{cases}
\label{eq:cameraD}
\end{equation}
\subsubsection{Probe visibility} Only the probes encompassing a geometry participates in the deferred shading. Thus, probes closer to a geometric surface are more important. Similarly, texels facing away from the surface are queried more often for shading. We express both quantities together in equation \ref{eq:ProbeVis}, where $p$ represents probe location and $t = trace(p, -\omega)$. The function \textit{trace} returns the distance of the nearest surface hit, and the scalar $s$ is the diagonal distance of a grid voxel.

\begin{equation}
\label{eq:ProbeVis} 
f_v(p, \omega) = e^{-2t/s}
\end{equation}

\subsubsection{Incoming radiance} We consider directions with high incoming radiance as more important. To identify those directions, we query the radiance along each probe octant and use it as a representative for incoming radiance. 

\begin{equation}
f_r(p, \omega) = \frac{min(r, \beta)}{\beta},
\end{equation}

where $r = lum(p, \omega)$. The function \textit{lum} returns the incoming luminance using direct illumination at the surface hit point. The parameter $\beta$ controls the dynamic range and we set $\beta = 5$.

\subsubsection{Probe visibility change \label{subsect:visChange}} Detection of dynamic geometry is crucial for increased resource allocation in regions affected by these changes. We detect dynamic geometry by computing a temporal gradient of probe visibility followed by a spatio-temporal smoothing operation.

\begin{equation}
f_0(p, \omega) = f_v^t(p, \omega) - f_v^{t-1}(p, \omega),
\label{eq:tempGradVis}
\end{equation}

where $f_v^t, f_v^{t-1}$ represent visibility in the current and last time step respectively. Equation \ref{eq:tempGradVis} implicitly states we keep the position and the direction fixed when measuring the time difference across frames to avoid noisy gradients. The gradient is passed through a temporal trigger ($Tr$) as: 

\begin{equation}
f_1(p, \omega) = Tr\left(f_0(p, \omega), \theta\right),
\end{equation}

\begin{figure}[t!]
\definecolor{OrangeYellow}{rgb}{0.929, 0.49, 0.192}
\begin{tikzpicture}
    \node[anchor=south west,inner sep=0] at (0,0){\includegraphics[width=13.93cm, trim={0cm 11.5cm 0.1cm 0cm},clip]{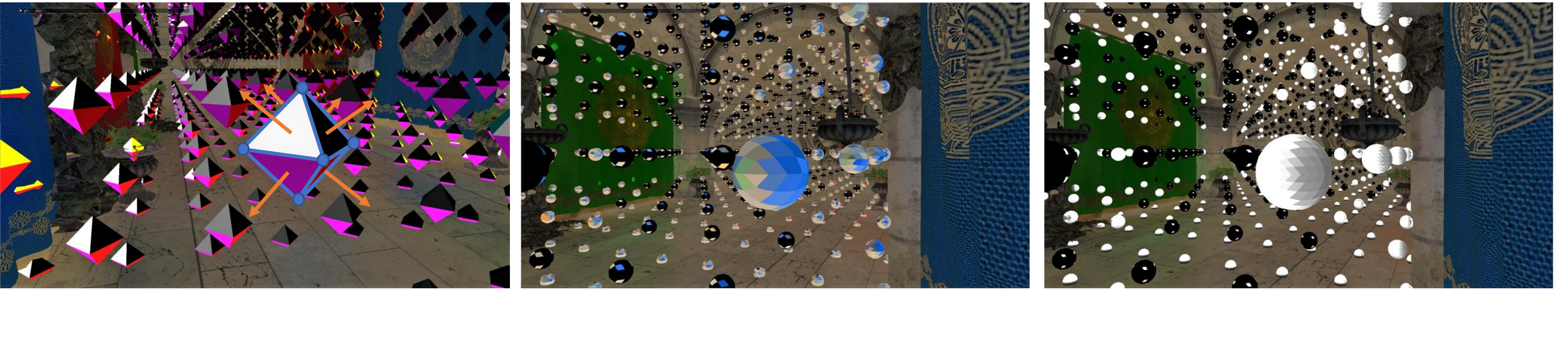}}; 
    \node[anchor=south west] at (0.0, 0.1) { \textcolor{black}{\small{\footnotesize a. Probes storing prior-information ($f$).} }};
    \node[anchor=south west] at (1.6, 0.85) { \textcolor{white}{\footnotesize octant-mapped}};
    \node[anchor=south west] at (1.9, 0.6) { \textcolor{white}{\footnotesize pilot-rays}};
    \node[anchor=south west] at (5.5, 0.1) { \textcolor{black}{\small{\footnotesize b. Irradiance cache ($\hat{g_r}$).} }};
    \node[anchor=south west] at (10.20, 0.1) { \textcolor{black}{\small{\footnotesize c. Visibility cache ($\hat{g_c}$).} }};
\end{tikzpicture}
\vspace{-25pt}
\caption{Figure showing various probe-mapped textures and LUT in our technique.}
\label{fig:probeStructure}
\end{figure}
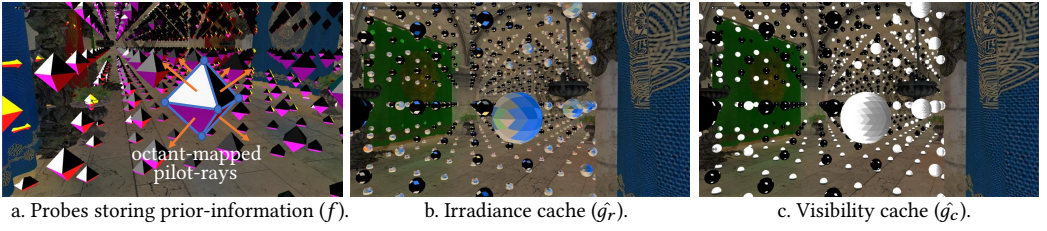

where $Tr$ converts a pulse in time to a decaying signal controlled by the parameter $\theta$ as shown in figure \ref{fig:TrCam}(a). For simplicity, we drop the time axis from the function $Tr$. The function minimizes temporal discontinuities, thus helping the Markov-chain to closely follow the target distribution ($h$) across frames. Finally, we perform a spatial convolution as follows:

\begin{equation}
f_{\Delta v}(p, \omega) = \sum\limits_{i, j}f_1(p - p_i, \omega - \omega_j).
\end{equation}

The convolution step smooths out uncertainties in a single texel and also serves as a weak predictor of possible locations of the dynamic geometry in the next frame. We use a $5 \times 5 \times 5$ and $3 \times 3$ convolution in space and direction,  respectively. 

\subsubsection{Probe radiance change} Similar to the previous section, we detect a change in radiosity using a temporal gradient of the probe radiance. We apply the same temporal trigger and spatial convolution operator as in the previous section. The corresponding equations are as follows:

\begin{equation}
\label{eq:tempGradIrr}
f_2(p, \omega) = f_r^t(p, \omega) - f_r^{t-1}(p, \omega),
\end{equation}

\begin{equation}
f_3(p, \omega) = Tr\left(f_2(p, \omega), \theta\right),
\end{equation}

\begin{equation}
\label{eq:irradChange}
f_{\Delta r}(p, \omega) = \sum\limits_{i, j}f_3(p - p_i, \omega - \omega_j).
\end{equation}

\subsection{Heuristics composition \label{sec:priorCompose}} 

Now that the individual heuristics are defined, as described in equation \ref{eq:guide}, we compose them for the static and dynamic cases as follows:

\begin{equation}
\begin{aligned} 
f_s(p, \omega) = \overbrace{f_c f_v}^{static},
\end{aligned}
\label{eq:priorStatic}
\end{equation}

\begin{equation}
\begin{aligned} 
f_d(p, \omega) = \overbrace{f_cf_v(f_{\Delta v} + \mu f_{\Delta r})}^{dynamic}.
\end{aligned}
\label{eq:priorDynamic}
\end{equation}

When the environment is static, we sample according to the camera and probe-to-surface distance heuristics denoted by $f_c$ and $f_v$ in equation \ref{eq:priorStatic}. In the dynamic case represented by equation \ref{eq:priorDynamic}, we modulate the changes in the environment by the static term $f_c f_v$. The modulation indicates we are more interested in changes close to the camera and geometric surfaces. The factor $\mu$ weighs the strength of change in geometry versus change in lighting. We use $\mu=2$ in all our experiments.

\subsection{Heuristics storage}

We store the quantities $f_v, f_s, f_d$ as a 6-10-10 bit encoded 32 bit integer at each octant of the probes. The remaining 6 bits are used for other flags. When querying the LUT/texture, we use a mapping function that maps the continuous position $p$ and direction $\omega$ to the corresponding texel in the LUT. We note that $f_c$ is implicitly defined, hence do not require additional storage. 

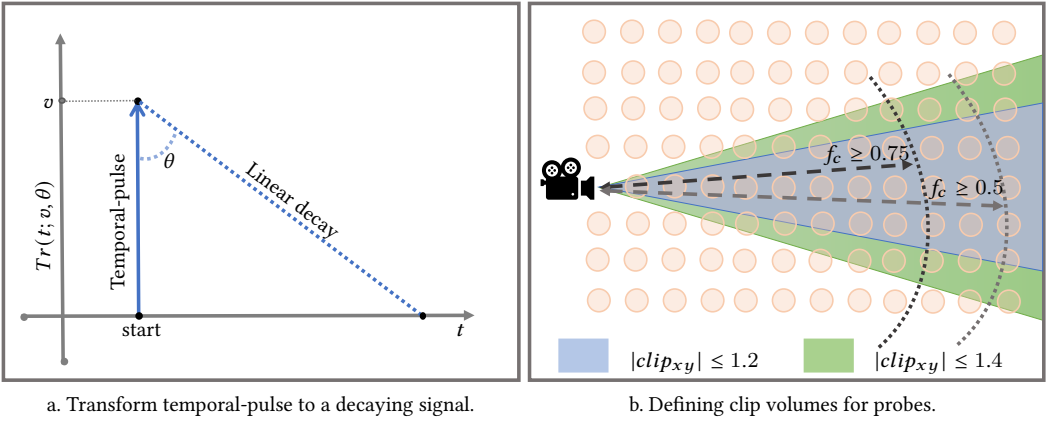
\begin{figure*}
\begin{tikzpicture}
    \node[anchor=south west,inner sep=0] at (0,0){\includegraphics[width=13.93cm, trim={0cm 9.5cm 10.5cm 0cm},clip]{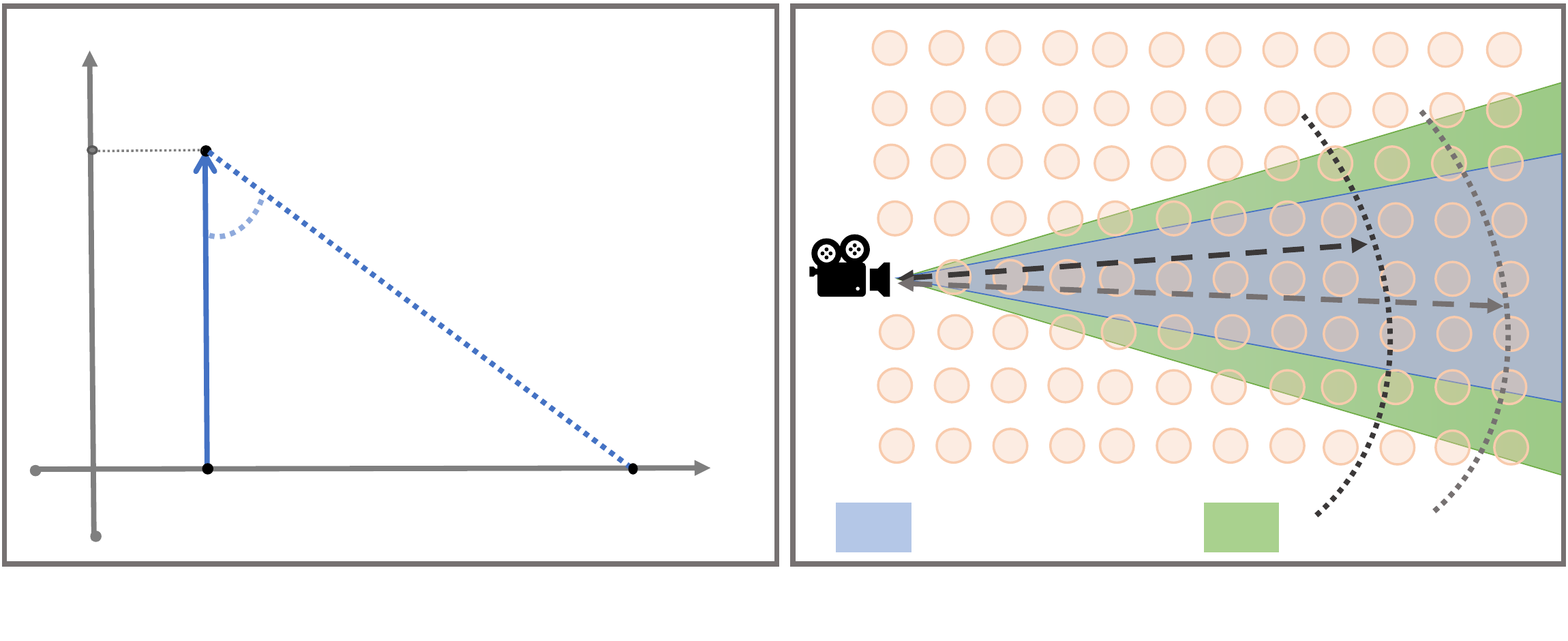}}; 
    
     \node[anchor=south west, rotate=90] at (0.85, 2.0) { \textcolor{black}{\footnotesize $Tr(t; v, \theta)$ }};
     \node[anchor=south west] at (0.45, 4.15) { \textcolor{black}{\footnotesize $v$}};
     \node[anchor=south west, rotate=90] at (1.85, 1.75) { \textcolor{black}{\footnotesize Temporal-pulse}};
     \node[anchor=south west] at (2.05, 3.35) { \textcolor{black}{\footnotesize $\theta$ }};
     \node[anchor=south west, rotate=323] at (3.05, 3.35) { \textcolor{black}{\footnotesize Linear decay }};
     \node[anchor=south west] at (1.5, 1.15) { \textcolor{black}{\footnotesize start} };
     \node[anchor=south west] at (5.95, 1.15) { \textcolor{black}{\footnotesize $t$ }};
    \node[anchor=south west] at (0.5, 0.1) { \textcolor{black}{\footnotesize a. Transform temporal-pulse to a decaying signal. }};
    
    \node[anchor=south west] at (10.85, 3.4) { \textcolor{black}{\footnotesize $f_c \geq 0.75$ }};
    \node[anchor=south west] at (12.25, 2.95) { \textcolor{black}{\footnotesize $f_c \geq 0.5$ }};
    \node[anchor=south west] at (8.25, 0.65) { \textcolor{black}{\footnotesize $|clip_{xy}| \leq 1.2$ }};
    \node[anchor=south west] at (12.25 - 0.75, 0.65) { \textcolor{black}{\footnotesize $|clip_{xy}| \leq 1.4$ }};
    \node[anchor=south west] at (8.25, 0.1) { \textcolor{black}{\footnotesize b. Defining clip volumes for probes.}};
    
\end{tikzpicture}
\vspace{-22pt}
\caption{Figure (a) shows the construction of temporal-trigger $Tr(v, \theta)$. In figure (b), we call the volume bounded by the blue frustum and black boundary as inner volume $V_{in}$. Similarly, outer volume $V_{out}$ is the volume bounded by green frustum and outer grey boundary. All probes in $V_{out}$ participate in the heuristic modelling, as described in section \ref{subsec:priorEfficient}. Probes inside the blue frustum participate in adaptive sampling as described in section \ref{subsect:samplingStaticDetails}, \ref{subsect:samplingDetails}. We set the probe state $N=16$ for all probes outside $V_{in}$ but inside $V_{out}$, refer section \ref{subsect:samplingStaticDetails}.}
\label{fig:TrCam}
\end{figure*}

\subsection{Improving construction efficiency \label{subsec:priorEfficient}}
The heuristics construction step is a potential bottleneck if we trace $8$ rays per probe for all probes in the scene. As such, we restrict the pilot-rays to the probes that are contained within an extended camera frustum as shown in figure \ref{fig:TrCam}(b). To maximize the efficiency of our algorithm, we further reuse the samples collected from the 8 pilot-rays to populate the irradiance ($\hat{g_r}$) and visibility ($\hat{g_c}$) caches. We change the ray-directions at alternate frames in an \textit{AABBCCDD...} pattern, improving the detection of temporally varying light-field surrounding the probes. We measure the time-delta (equation \ref{eq:tempGradVis} and \ref{eq:tempGradIrr}) between two frames with identical set of ray-queries, avoiding noisy gradients. However, this effectively halves the detection frequency (\textit{frame-rate / 2}) but improves the spatial awareness. We use a stratified-random ray-direction such that there is always one ray per octant. We update the irradiance and visibility cache at each alternate frame.

\subsection{Probe irradiance cache \label{subsect:probeIrCach}}

As shown in figure \ref{fig:probeStructure}(b), the irradiance cache ($\hat{g_r}$) is represented as a uniform probe grid in space where each probe stores the surrounding diffuse irradiance at a $8 \times 8$ texel resolution using a spherical mapping. At each texel, we store the irradiance in a custom RGB encoding with 9-9-8 bits for the three channels. The remaining 6 bits (out of 32bit) store the sample accumulation count (N), used for computing the moving average (see algorithm \ref{alg:MovingAvgUpdate}) of a sample stream in time. We take several considerations into account for the choice of our encoding. Our encoding should be bandwidth efficient and must support atomic updates on a commodity GPU. We found both DX12 and GLSL supports atomic operations on 32 bit integers. Finally, our encoding must faithfully encode intensities beyond the standard definition. We apply a non-linear color compression across the three color channels, $i \in [0..2]$ as shown in the equation below.

\begin{equation}
    u_i = \frac{min\left(ln(\gamma \cdot v_i + 1), \beta\right)}{\beta}.
    \label{eq:rgbEnc}
\end{equation}

We apply an inverse transform $\left(exp(\beta \cdot u_i) - 1\right)/\gamma$ while decoding where $\beta=5$ and $\gamma=15$. More details regarding our choice of compression scheme is provided in appendix \ref{sec:probeCompression} and figure \ref{fig:rgbEnc}.

\subsection{Probe visibility cache \label{subsect:probeVisCach}}

As shown in figure \ref{fig:probeStructure}(c), texels in the visibility probes store the mean distances and mean squared distances to the nearest geometry at 16x16 texel resolution. We call this $\hat{g_c}$ - our visibility cache. Each texel stores the two channels with 13 bits of precision each while the rest 6 bits are used for sample accumulation count. We normalize the distances with probe cage diagonal length. Similar to irradiance cache, we apply a logarithmic encoding as per equation \ref{eq:rgbEnc} for efficient use of available precision. We use $(\beta, \gamma)$ values of $(5, 15)$ and $(8, 20)$ for the linear and squared channels respectively.

\subsection{Temporal sample accumulation mecahnism} We use a moving-average accumulation to store the samples in the irradiance and visibility caches. In the algorithm \ref{alg:MovingAvgUpdate}, we have two parameters $N$ and $N_{max}$ to control the moving-average accumulation. As we start accumulating samples, $N$ is incremented and the algorithm performs like a true moving average. However, as $N$ approaches $N_{max} - 1$, the algorithm switches to an exponential moving average form with hysteresis $(N_{max} - 1) / N_{max}$. Also, note that when the value of $N$ is low, the cache updates itself quickly, but the stored values may be noisy. As $N$ increases, the new samples are weighed less in their contribution to the cache. We exploit these parameters to control the learning rate and noise in the static and dynamic cases as discussed in the following sections.  

\begin{algorithm}[t!]
\caption{Moving Average algorithm}\label{alg:MovingAvgUpdate}
\DontPrintSemicolon
\KwInput{$x$: Update location, $v$ : New sample, $N_{max}$ : Max sample count}
\KwOutput{$V$ : Updated value, $N$:  Sample count}
\SetKwFunction{MAV}{MovingAvgUpdate}
\SetKwProg{Fn}{function}{:}{}
  \Fn{\MAV{$x$, $v$, $N_{max}$}}
  {
    $n \gets \hat{g}[x].N$ \tcp*{Cumulative sample count}
    $o \gets \hat{g}[x].V$ \tcp*{Cumulative value}
    $V \gets \frac{v}{n + 1} + \frac{n \cdot o}{n + 1}$ \tcp*{Update cumulative value}
    $N \gets min(n + 1, N_{max})$ \tcp*{Increment sample count}
    \KwRet $V, N$\;
  }
\end{algorithm}

\subsection{Adaptive sampling - static\label{subsect:samplingStaticDetails}}

We split our adaptive sampling strategy into two stages - static and dynamic. We have two separate Markov-chain sets, each focusing on different aspects of capturing the surrounding light-field. While the static chain focuses more on the accuracy, the dynamic chain is tuned for capturing the transient responses. We discuss the dynamic chain in detail in the next section. 

\textcolor{black}{We set up equation \ref{eq:DDGIBays} as - $h = exp(min(\hat{g}_r/f_s, 1))\cdot f_s$. The feedback from irradiance cache $\hat{g}_r$ is obtained from the previous frame and from a higher mip-level (also used in deferred shader). The lowest mip-level $\hat{g}_r$ is continuously updated and thus avoided as feedback due to possible violation of stationarity condition within a frame.}     
We use the Metropolis sampling, algorithm \ref{alg:MarkovChainMT}, to generate the samples $x_i \equiv (p_i, \omega_i)$. As summarized in the algorithm, \ref{alg:ApproxAlgo}, we use the samples to evaluate the continuous light field $g_r$, which involves tracing a ray originating at $p_i$ along the direction $\omega_i$. We trace an additional shadow-ray per sample to compute the visibility in the opposite direction ($-\omega_i$) as the probe queries in the deferred shader for visibility is exactly $180^\circ$ out of phase w.r.t irradiance. Next we store the irradiance and visibility values in the irradiance ($\hat{g_r}$) and visibility ($\hat{g_c}$) caches using an atomic update rule as presented in the algorithm \ref{alg:atmMvAvgUpd}. Atomic updates are required as multiple invocations of the chain may update the same location in the irradiance and visibility caches. Figure \ref{fig:Method} summarizes the overall idea.

We set the random walk step size, denoted by $\sigma \in R^5$ in algorithm \ref{alg:MetropolisStep}, proportional to the size of discretization in the irradiance and visibility cache. Thus positional step size is proportional to the size of a voxel in the probe grid, while angular step size is roughly $\sqrt{\pi/256}$. Due to the small step size, texels in the cache may accumulate more than one sample per texel, thereby accumulating a large sample count over time. We also note that our cache behaves like a true moving average between sample count $N=0 \text{ to } 64$, which also contributes to better accuracy.

\begin{algorithm}[t!]
\caption{Atomic moving average algorithm}\label{alg:atmMvAvgUpd}
\DontPrintSemicolon
\KwInput{$x$: Update location, $v$ : New update value}
\KwOutput{Update $\hat{g}[x]$}
\SetKwFunction{AMAV}{AtomicMovingAvg}
\SetKwProg{Fn}{function}{:}{}
  \Fn{\AMAV{$x$, $v$}}
  { 
    current $\gets \hat{g}[x]$\;
    \tcc{Repeat until destination value stops changing}
    \KwDo{current $\neq$ expected}
    {
        expected $\gets$ current\;
        next $\gets$ MovingAvgUpdate$(x, v, 64)$\;
        InterlockedCompareExchange($\hat{g}[x]$, expected, next, current) \tcp*{Refer HLSL}
    }
 }
\end{algorithm}

The static adaptive samples are useful for improving convergence in a static scene and for slow changes that are undetected during prior construction. For example, slow changes in lighting such as day-night cycles in games. We lower the hysteresis by setting $N=16$ for all probes in the region $\{V_{out}\} - \{V_{in}\}$ in figure \ref{fig:TrCam}(b). This enables the probe to quickly catch-up to the most recent values.   

\subsection{Adaptive sampling - dynamic\label{subsect:samplingDetails}}

We run a second set of Markov-chain when dynamic content is detected in the scene. When there are dynamic elements, especially moving geometry, we run into two main issues. The generated samples are not well distributed in the region of interest i.e. the areas where time varying changes are present. When the step size is small, the chain cannot track the target distribution fast enough to generate samples from the target, causing the samples to lag the moving target distribution. The second problem is noise due to multi-sampling of the irradiance texel. Potentially, this can be solved by increasing the hysteresis to improve temporal sample reuse. However, the reduced noise comes at the cost of introducing objectionable temporal blur. 

We solve the first issue by increasing the chain step size and by coarsening the target function ($f_d$). Practically, this amounts to grouping the heuristics-probes into virtual proxies. In our case, a virtual proxy represents a group the $3 \times 3 \times 3$ probes. This virtual probe has 8 directions and each direction represents an axis-aligned octant. The value of a texel of the virtual probe is the \textit{max} of all 27 probes it represents along the corresponding direction. We also drop the sampled evidence by setting $\alpha =0$ in equation \ref{eq:DDGIBays}, as the stale irradiance cache ($\hat{g_r}$) provide little useful information for sampling a time varying region. The chain step size is 3x, and 6x larger for position and directions, respectively w.r.t the static case.

Since each sample from the coarse chain represents an entire octant, we trace 64 rays for the octant for all underlying 3x3x3 probes in the group. We make the tracing step more efficient by culling probes that are not used in deferred shading. The scheduling of ray-direction is deterministic, passing through the center of a texel in the irradiance cache ($\hat{g_r}$). This solves the problem of sampling noise and also affords the opportunity to simplify the atomic updates. Since the rays are not random, we do not benefit from multiple shader invocations updating the same octant. As such, the first invocation to update the octant marks (atomically) it updated such that other invocations do not repeat the same work move to the next.

We run the dynamic sampling after the static sampling step. During static sampling, if a probe has non-zero dynamic component($f_d > 0$), we quantize the ray directions to go through the irradiance/visibility cache texel center to avoid injecting sampling noise in the texels.

\section{Results and comparisons \label{sec:results}}

\begin{table*}[t!]
\centering
\caption{\label{tab:probeDetails} Table showing probe grid details for various scenes used in our technique.}
\vspace{-8pt}
\begin{tabular}{|c|c|c|c|c|}
\hline
Scene & Probe Grid & \begin{tabular}[c]{@{}c@{}}Probe spacing\\ (in meters)\end{tabular} & \begin{tabular}[c]{@{}c@{}}Irradiance ($\hat{g_r}$)\\ Cache Resolution\end{tabular} & \begin{tabular}[c]{@{}c@{}}Visibility ($\hat{g_c}$) \\ Cache Resolution\end{tabular} \\ \hline
Bistro - Exterior & $192\times64\times192$ & $0.5\times0.5\times0.5$ & $8\times8$ & $16\times16$ \\ \hline
Sponza - Diffuse &  $192\times64\times192$ & $0.5\times0.5\times0.5$ & $8\times8$ & $16\times16$ \\ \hline
Sponza - Glossy &  $192\times64\times192$ & $0.1\times0.1\times0.1$ & $16\times16$ & $16\times16$ \\ \hline
\end{tabular}
\end{table*}

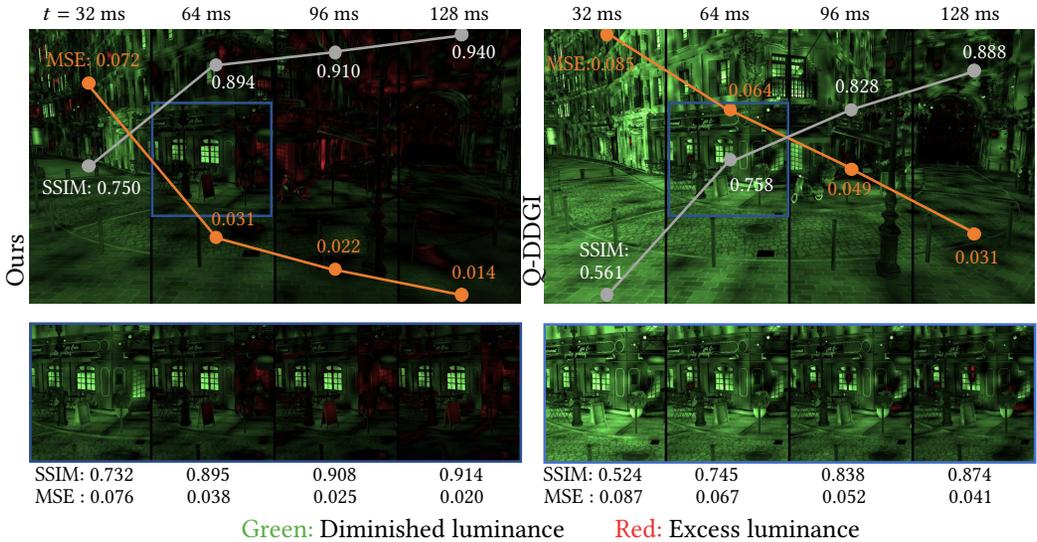
\begin{figure}[b!]
\definecolor{OrangeYellow}{rgb}{0.929, 0.49, 0.192}
\definecolor{SomewhatGreen}{rgb}{0.356, 0.678, 0.278}
\definecolor{PinkishRed}{rgb}{0.929, 0.192, 0.192}
\begin{tikzpicture}
    \node[anchor=south west,inner sep=0] at (0,0){\includegraphics[width=13.5cm, trim={0cm 5.5cm 2.35cm 0cm},clip]{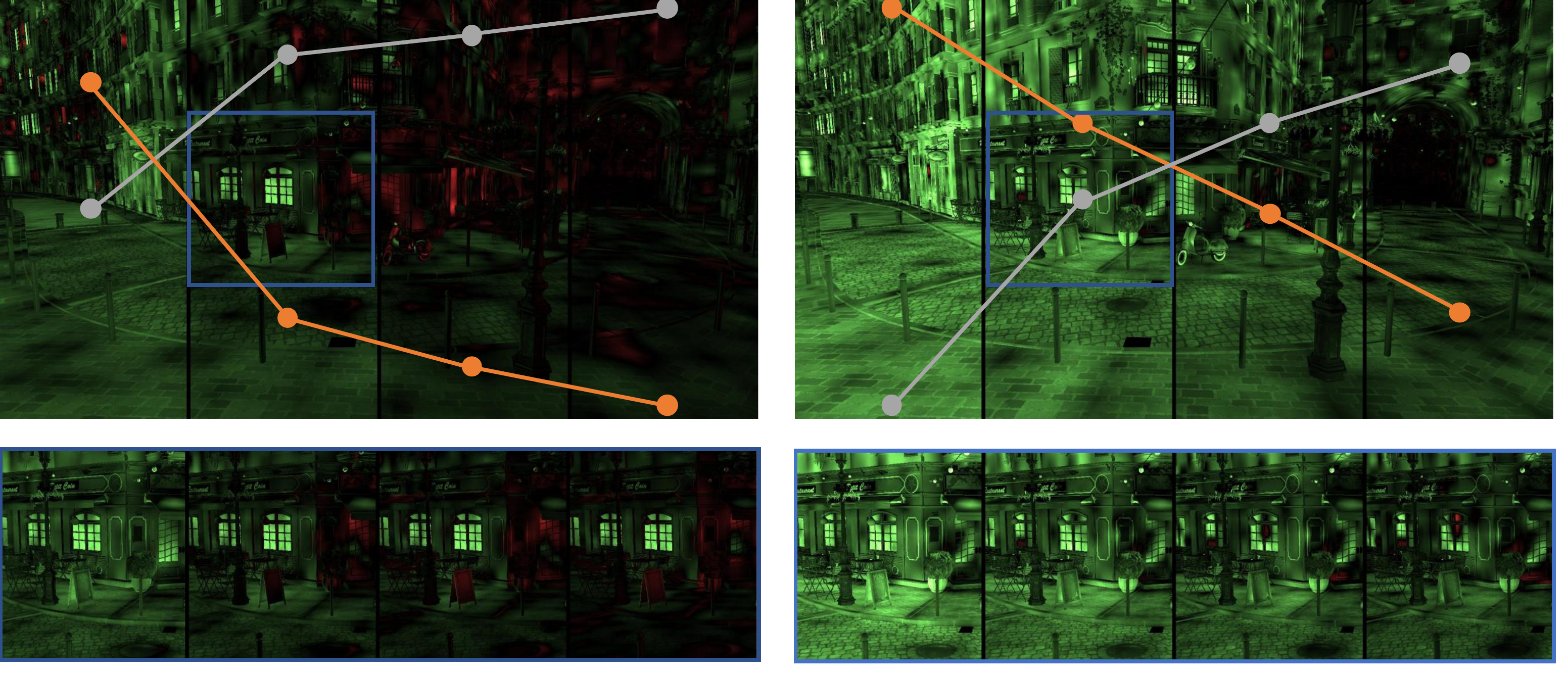}}; 
     \node[anchor=south west] at (0.05,5.8) { \textcolor{black}{\footnotesize $t= \text{32 ms}$} };
     \node[anchor=south west] at (1.9,5.8) { \textcolor{black}{\footnotesize 64 ms }};
     \node[anchor=south west] at (3.6,5.8) { \textcolor{black}{\footnotesize 96 ms} };
     \node[anchor=south west] at (5.2,5.8) { \textcolor{black}{\footnotesize 128 ms} };
     \node[anchor=south west] at (7.1,5.8) { \textcolor{black}{\footnotesize 32 ms} };
     \node[anchor=south west] at (8.8,5.8) { \textcolor{black}{\footnotesize 64 ms }};
     \node[anchor=south west] at (10.4,5.8) { \textcolor{black}{\footnotesize 96 ms} };
     \node[anchor=south west] at (12.0,5.8) { \textcolor{black}{\footnotesize 128 ms} };
     \node[anchor=south west, rotate=90] at (0.05,2.3) { \textcolor{black}{Ours}};
     \node[anchor=south west, rotate=90] at (7.0,2.3) { \textcolor{black}{ Q-DDGI }};
     \node[anchor=south west] at (0.1,5.2){\textcolor{OrangeYellow}{\footnotesize MSE: 0.072}};
     \node[anchor=south west] at (2.3,3.1){\textcolor{OrangeYellow}{\footnotesize 0.031}};
     \node[anchor=south west] at (3.7,2.75){\textcolor{OrangeYellow}{\footnotesize 0.022}};
     \node[anchor=south west] at (5.5,2.4){\textcolor{OrangeYellow}{\footnotesize 0.014}};
      \node[anchor=south west] at (0.05,3.52) { \textcolor{white}{\footnotesize SSIM: 0.750} };
      \node[anchor=south west] at (2.3,4.9) { \textcolor{white}{\footnotesize 0.894} };
      \node[anchor=south west] at (3.7,5.05) { \textcolor{white}{\footnotesize 0.910} };
      \node[anchor=south west] at (5.5,5.3) { \textcolor{white}{\footnotesize 0.940} };
    \node[anchor=south west] at (6.75,5.15){\textcolor{OrangeYellow}{\footnotesize MSE:}};
    \node[anchor=south west] at (7.35,5.15){\textcolor{OrangeYellow}{\footnotesize 0.085}};
     \node[anchor=south west] at (9.18,4.8){\textcolor{OrangeYellow}{\footnotesize 0.064}};
     \node[anchor=south west] at (10.5,3.5){\textcolor{OrangeYellow}{\footnotesize 0.049}};
     \node[anchor=south west] at (12.2,2.6){\textcolor{OrangeYellow}{\footnotesize 0.031}};
      \node[anchor=south west] at (7.2,2.7) { \textcolor{white}{\footnotesize SSIM:}};
      \node[anchor=south west] at (7.2,2.4) { \textcolor{white}{\footnotesize 0.561}};
      \node[anchor=south west] at (9.2,3.55) { \textcolor{white}{\footnotesize 0.758}};
      \node[anchor=south west] at (10.6,4.85) { \textcolor{white}{\footnotesize 0.828} };
      \node[anchor=south west] at (12.3,5.3) { \textcolor{white}{\footnotesize 0.888} };
      \node[anchor=south west] at (-0.05,-0.25) { \textcolor{black}{\footnotesize{SSIM: 0.732\hspace{7mm}0.895\hspace{11mm}0.908\hspace{11mm}0.914\hspace{8mm}SSIM: 0.524\hspace{7mm}0.745\hspace{11mm}0.838\hspace{11mm}0.874}}};
      \node[anchor=south west] at (-0.05,-0.55) { \textcolor{black}{\footnotesize{MSE : 0.076\hspace{7mm}0.038\hspace{11mm}0.025\hspace{11mm}0.020\hspace{8mm}MSE : 0.087\hspace{7mm}0.067\hspace{11mm}0.052\hspace{11mm}0.041}}};
      \node[anchor=south west] at (2.7,-1.0) { \textcolor{black}{\textcolor{SomewhatGreen}{Green:} Diminished luminance \hspace{5mm} \textcolor{PinkishRed}{Red:} Excess luminance} };
\end{tikzpicture}
\vspace{-18pt}
\caption{Comparing the convergence of our technique over time on a static \textsc{Bistro Exterior} scene. The figure demonstrates the effectiveness of our static adaptive sampling step. The two rows measure the difference in luminance w.r.t reference and highlight the error in red and green color.}
\label{fig:bistroExtStatic}
\end{figure}

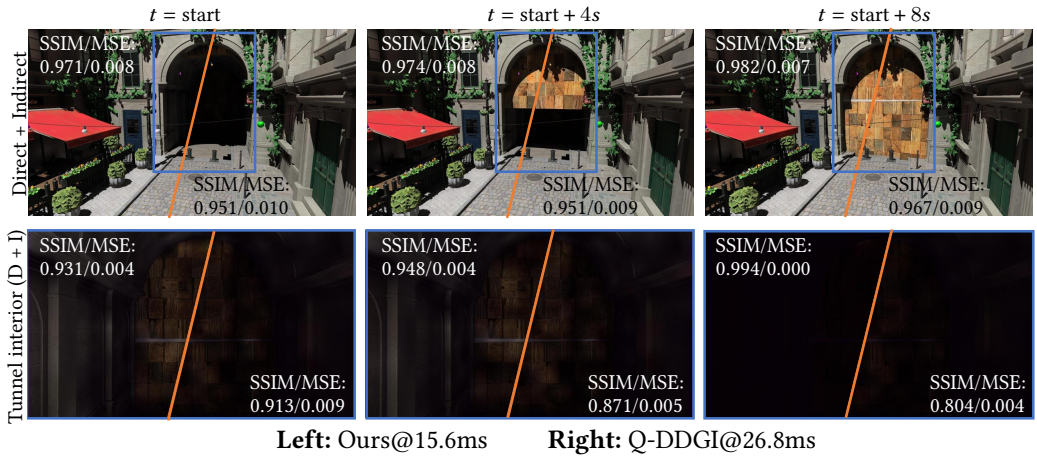
\begin{figure*}[t!]
\begin{tikzpicture}
    \node[anchor=south west,inner sep=0] at (0,0){\includegraphics[width=13.5cm, trim={0cm 6.1cm 2.35cm 0cm},clip]{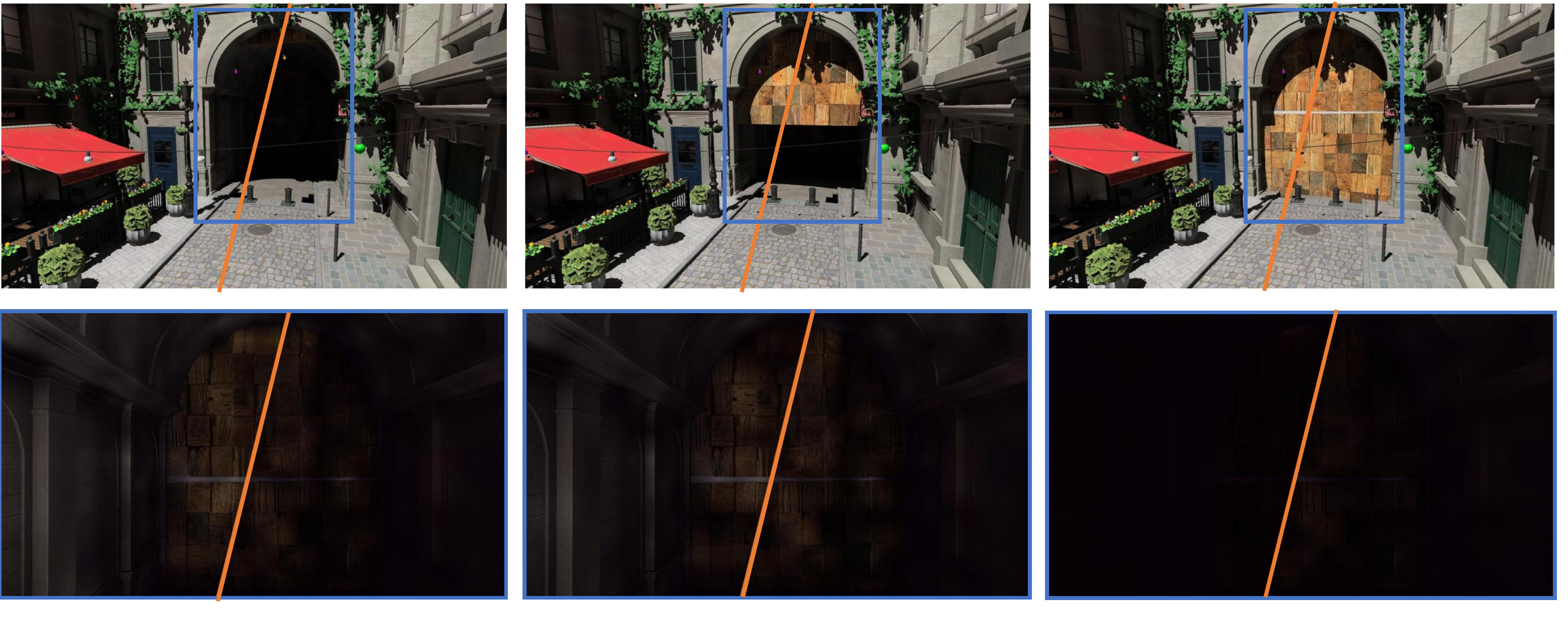}}; 
     \node[anchor=south west] at (1.5,5.5) { \textcolor{black}{\footnotesize $t= \text{start}$} };
     \node[anchor=south west] at (6,5.5) { \textcolor{black}{\footnotesize $t= \text{start} + 4s$} };
     \node[anchor=south west] at (10.4,5.5) { \textcolor{black}{\footnotesize $t= \text{start} + 8s$} };
    \node[anchor=south west, rotate=90] at (0.12,3.2) { \textcolor{black}{\footnotesize Direct + Indirect} };
    \node[anchor=south west, rotate=90] at (0.12,0.2) { \textcolor{black}{\footnotesize Tunnel interior (D + I)} };
    \node[anchor=south west] at (0.05,1.8+3.3) { \textcolor{white}{\footnotesize SSIM/MSE:} };
    \node[anchor=south west] at (0.05,1.8+3.0) { \textcolor{white}{\footnotesize 0.971/0.008} };
    \node[anchor=south west] at (4.6,1.8+3.3) { \textcolor{white}{\footnotesize SSIM/MSE:} };
    \node[anchor=south west] at (4.6,1.8+3.0) { \textcolor{white}{\footnotesize 0.974/0.008} };
    \node[anchor=south west] at (9.05,1.8+3.3) { \textcolor{white}{\footnotesize SSIM/MSE:} };
    \node[anchor=south west] at (9.05,1.8+3.0) { \textcolor{white}{\footnotesize 0.982/0.007} };
    \node[anchor=south west] at (0.45 + 1.65,0.5+2.75) { \textcolor{black}{\footnotesize SSIM/MSE:} };
    \node[anchor=south west] at (0.45 + 1.65,0.5+2.45) { \textcolor{black}{\footnotesize  0.951/0.010} };
    \node[anchor=south west] at (3.95 + 2.8,0.5+2.75) { \textcolor{black}{\footnotesize SSIM/MSE:} };
    \node[anchor=south west] at (3.95 + 2.8,0.5+2.45) { \textcolor{black}{\footnotesize 0.951/0.009} };
    \node[anchor=south west] at (8.7 + 2.65,0.5+2.75) { \textcolor{black}{\footnotesize SSIM/MSE:} };
    \node[anchor=south west] at (8.7 + 2.65,0.5+2.45) { \textcolor{black}{\footnotesize 0.967/0.009} };
    \node[anchor=south west] at (0.05,2.45) { \textcolor{white}{\footnotesize SSIM/MSE:} };
    \node[anchor=south west] at (0.05,2.15) { \textcolor{white}{\footnotesize 0.931/0.004} };
    \node[anchor=south west] at (4.6,2.45) { \textcolor{white}{\footnotesize SSIM/MSE: } };
    \node[anchor=south west] at (4.6,2.15) { \textcolor{white}{\footnotesize 0.948/0.004} };
    \node[anchor=south west] at (9.05,2.45) { \textcolor{white}{\footnotesize SSIM/MSE: } };
    \node[anchor=south west] at (9.05,2.15) { \textcolor{white}{\footnotesize  0.994/0.000} };
    \node[anchor=south west] at (0.45 + 2.4,0.65) { \textcolor{white}{\footnotesize SSIM/MSE:} };
    \node[anchor=south west] at (0.45 + 2.4,0.35) { \textcolor{white}{\footnotesize  0.913/0.009} };
    \node[anchor=south west] at (3.95 + 3.4,0.65) { \textcolor{white}{\footnotesize SSIM/MSE: } };
    \node[anchor=south west] at (3.95 + 3.4,0.35) { \textcolor{white}{\footnotesize  0.871/0.005} };
    \node[anchor=south west] at (8.7 + 3.2,0.65) { \textcolor{white}{\footnotesize SSIM/MSE:} };
    \node[anchor=south west] at (8.7 + 3.2,0.35) { \textcolor{white}{\footnotesize  0.804/0.004} };
    \node[anchor=south west] at (3.2,-0.2) { \textbf{Left:} Ours@15.6ms\hspace{8mm}\textbf{Right:} Q-DDGI@26.8ms};
\end{tikzpicture}
\vspace{-20pt}
\caption{Our technique compared with Q-DDGI on a modified \textsc{Bistro Exterior} scene augmented with a moving door. The scene has $192\times64\times192$ probes and shows the convergence of the two techniques near a dynamic area in the scene. The second row shows the changes inside the tunnel as the door closes over time. Our technique is better able to allocate the resources closer to the dynamic areas resulting in faster convergence and higher performance.}
\label{fig:bistroExtClose}
\end{figure*}

\begin{figure*}[t!]
\definecolor{SomewhatGreen}{rgb}{0.356, 0.678, 0.278}
\definecolor{PinkishRed}{rgb}{0.929, 0.192, 0.192}
\begin{tikzpicture}
    \node[anchor=south west,inner sep=0] at (0,0){\includegraphics[width=13.5cm, trim={0cm 15.5cm 2.35cm 0cm},clip]{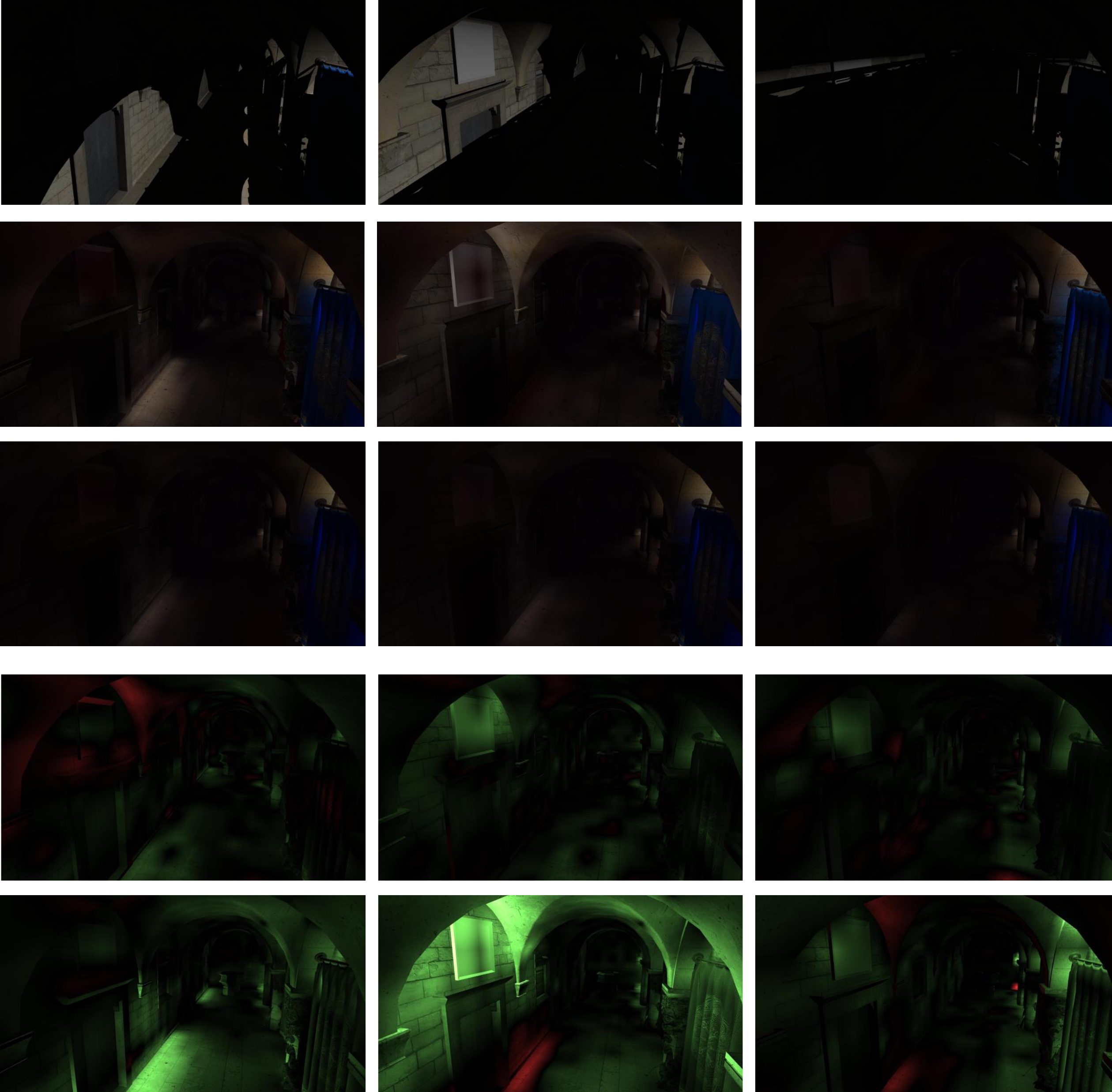}}; 
     \node[anchor=south west] at (1.3,3.7+16.9-7.3) { \textcolor{black}{\footnotesize $t= \text{start}$} };
     \node[anchor=south west] at (6,3.7+16.9-7.3) { \textcolor{black}{\footnotesize $t= \text{start} + 2s$} };
     \node[anchor=south west] at (10.3,3.7+16.9-7.3) { \textcolor{black}{\footnotesize $t= \text{start} + 3s$} };
    \node[anchor=south west, rotate=90] at (0.12,2.3 + 16-7) { \textcolor{black}{\footnotesize Direct only} };
    \node[anchor=south west, rotate=90] at (0.1,0.8 + 15-7.5) { \textcolor{black}{\footnotesize Indirect - Ours} };
    \node[anchor=south west, rotate=90] at (0.1,0.5 + 12.0-7) { \textcolor{black}{\footnotesize Indirect - Q-DDGI} };
    
    \node[anchor=south west] at (0.1, 17.3-7.15) { \textcolor{white}{\footnotesize SSIM/MSE: 0.920/0.005} };
    \node[anchor=south west] at (4.6, 17.3-7.15) { \textcolor{white}{\footnotesize SSIM/MSE: 0.966/0.006} };
    \node[anchor=south west] at (9.15, 17.3-7.15) { \textcolor{white}{\footnotesize SSIM/MSE:0.957/0.007} };
    
    \node[anchor=south west] at (0.1, 14.45-7.0) { \textcolor{white}{\footnotesize SSIM/MSE: 0.868/0.012} };
    \node[anchor=south west] at (4.6, 14.45-7.0) { \textcolor{white}{\footnotesize SSIM/MSE: 0.763/0.023} };
    \node[anchor=south west] at (9.15, 14.45-7.0) { \textcolor{white}{\footnotesize SSIM/MSE: 0.794/0.021} };
    
    \node[anchor=south west] at (3.9,7.1+4.9-6.975) { \textcolor{black}{Difference in luminance w.r.t reference} };
    
    \node[anchor=south west, rotate=90] at (0.1,0.8 + 16-7-6.65) { \textcolor{black}{\footnotesize Error - Ours} };
    \node[anchor=south west, rotate=90] at (0.1,0.5 + 13.3-7-6.65) { \textcolor{black}{\footnotesize Error - Q-DDGI} };
    \node[anchor=south west] at (3.1,-0.5) { \textcolor{black}{\textcolor{SomewhatGreen}{Green:} Diminished luminance \hspace{5mm} \textcolor{PinkishRed}{Red:} Excess luminance} };
    \node[anchor=south west] at (4.2,-1.0) { Ours@9.86ms\hspace{8mm}Q-DDGI@13.2ms};
\end{tikzpicture}
\vspace{-20pt}
\caption{Figure comparing the convergence of our technique under dynamic lighting controlled by the direct component shown in the first row. The last two rows measure the difference in luminance w.r.t reference and highlight the error in red and green color.}
\label{fig:spnzaDynEmit}
\end{figure*}
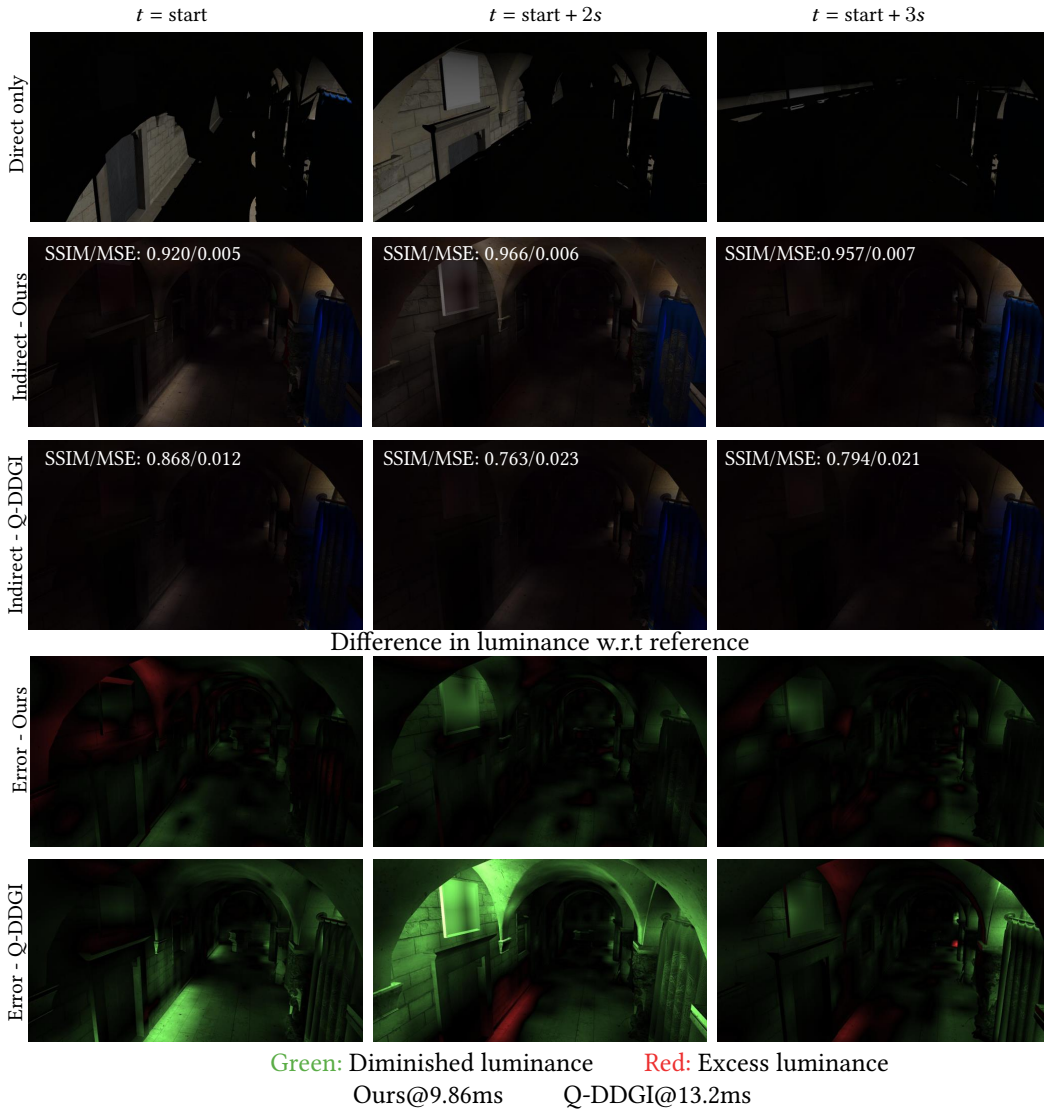

\begin{figure*}
\definecolor{SomewhatGreen}{rgb}{0.356, 0.678, 0.278}
\definecolor{PinkishRed}{rgb}{0.929, 0.192, 0.192}
\begin{tikzpicture}
    \node[anchor=south west,inner sep=0] at (0,0){\includegraphics[width=13.5cm, trim={0cm 11.5cm 2.1cm 0cm},clip]{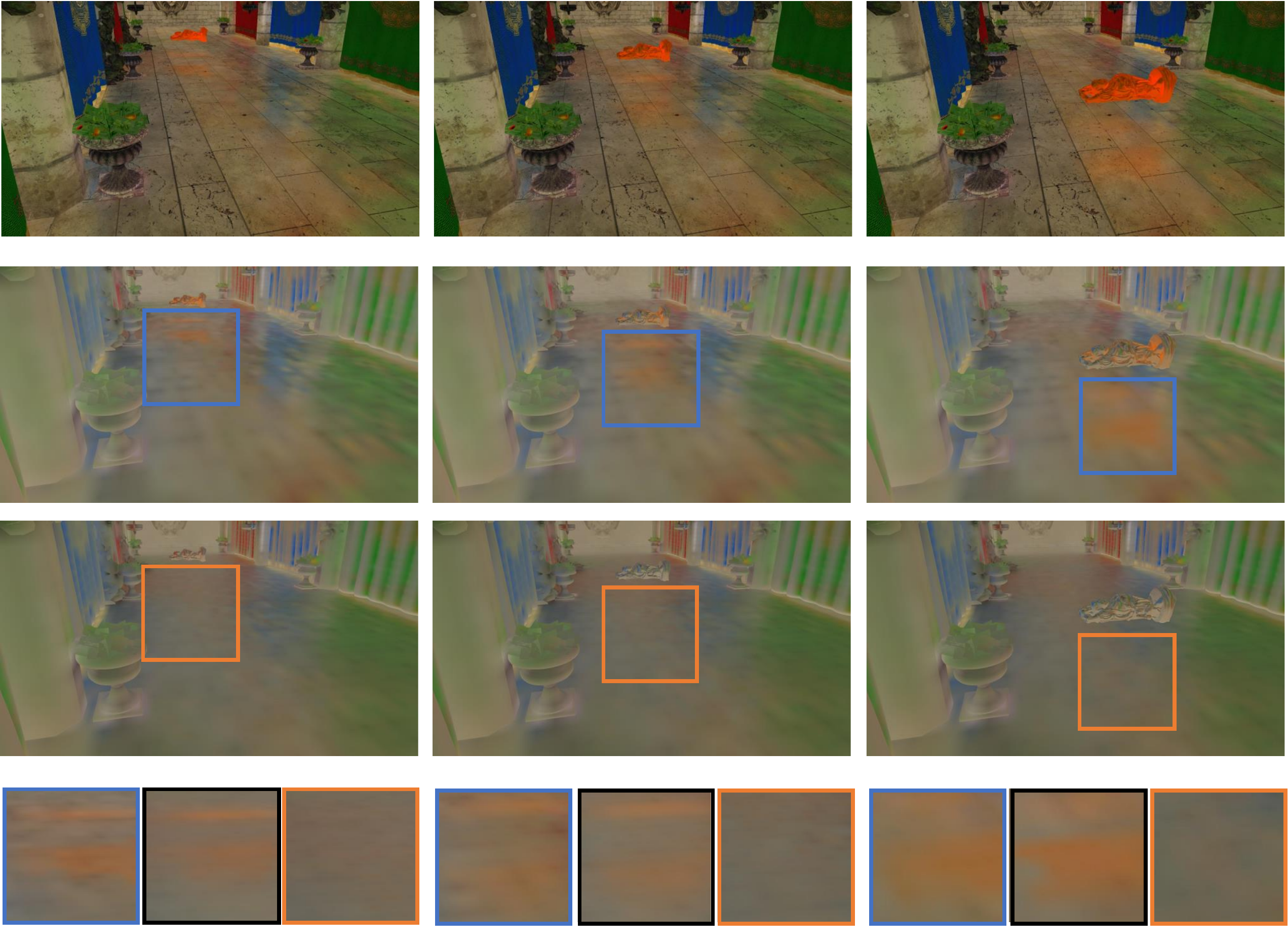}}; 
     \node[anchor=south west] at (1.4, 9.9) { \textcolor{black}{\footnotesize $t= \text{start}$} };
     \node[anchor=south west] at (5.8, 9.9) { \textcolor{black}{\footnotesize $t= \text{start} + 10s$} };
     \node[anchor=south west] at (10.25, 9.9) { \textcolor{black}{\footnotesize $t= \text{start} + 15s$} };
     \node[anchor=south west, rotate=90] at (0.1, 7.3) { \textcolor{black}{\footnotesize Glossy indirect - Ours} };
     
     \node[anchor=south west] at (4.4, 7.0) { \textcolor{black}{Glossy indirect without texture}};
     
     \node[anchor=south west, rotate=90] at (0.1, 5.5) { \textcolor{black}{\footnotesize Ours} };
      \node[anchor=south west, rotate=90] at (0.1,2.8) { \textcolor{black}{\footnotesize Q-DDGI} };
      
       \node[anchor=south west] at (0.03, 6.75) { \textcolor{black}{\footnotesize{ SSIM/MSE:0.996/0.017} }};
       \node[anchor=south west] at (4.5, 6.75) { \textcolor{black}{\footnotesize{ SSIM/MSE:0.995/0.019} }};
       \node[anchor=south west] at (9.1, 6.75) { \textcolor{black}{\footnotesize{ SSIM/MSE:0.992/0.018} }};
       
       \node[anchor=south west] at (0.03, 4.07) { \textcolor{black}{\footnotesize{ SSIM/MSE:0.990/0.019} }};
       \node[anchor=south west] at (4.5, 4.07) { \textcolor{black}{\footnotesize{ SSIM/MSE:0.989/0.020} }};
       \node[anchor=south west] at (9.1, 4.07) { \textcolor{black}{\footnotesize{ SSIM/MSE:0.987/0.028} }};
      
      \node[anchor=south west] at (0.05, 1.57) { \textcolor{black}{\footnotesize{ \hspace{2mm}Ours\hspace{6mm}Reference\hspace{5mm}Q-DDGI
      \hspace{8mm}Ours\hspace{6mm}Reference\hspace{5mm}Q-DDGI
      \hspace{6mm}Ours\hspace{6mm}Reference\hspace{5mm}Q-DDGI}}};
      
      \node[anchor=south west] at (-0.05, -0.15) { \textcolor{black}{\footnotesize{ SSIM: 0.979\hspace{20mm}0.981
      \hspace{6mm}SSIM: 0.995\hspace{20mm}0.993
      \hspace{6mm}SSIM: 0.998\hspace{20mm}0.992}}};
      
      \node[anchor=south west] at (-0.05, -0.5) { \textcolor{black}{\footnotesize{ MSE : 0.013\hspace{20mm}0.028
      \hspace{6mm}MSE : 0.016\hspace{20mm}0.034
      \hspace{6mm}MSE : 0.014\hspace{20mm}0.058}}};
      
      \node[anchor=south west] at (4.2,-1.0) { Ours@19.7ms\hspace{8mm}Q-DDGI@31.6ms};
\end{tikzpicture}
\vspace{-20pt}
\caption{Figure comparing glossy indirect reflection on a scene lit by ambient lighting. The scene tests transient response due to the moving \textsc{Buddha} geometry over a glossy floor.}
\label{fig:spnzaDynBuddha}
\end{figure*}

We compare our results with Q-DDGI and a reference probe-based implementation in different scenarios - static scene (fig. \ref{fig:bistroExtStatic}), dynamic geometry (fig. \ref{fig:teaser}, \ref{fig:bistroExtClose}, \ref{fig:spnzaDynBuddha}), and dynamic lighting (fig. \ref{fig:spnzaDynEmit}).

\textit{Q-DDGI:} Quantized-DDGI or Q-DDGI is a performance enhanced extension of original DDGI \citep{DDGI19}, achieved without major modifications to the base algorithm. Q-DDGI is equipped with a more compact irradiance and visibility cache representation that closely resembles ours. See table \ref{tab:cacheDetails}. We also enable camera-frustum culling of probes in Q-DDGI as described in section \ref{subsec:priorEfficient} and figure \ref{fig:TrCam}. These modifications allow Q-DDGI to have similar performance (table \ref{tab:performanceDetails}) at same probe count (table \ref{tab:probeDetails}) as ours across different scenes. We believe these modifications make our comparisons more fair. We use 32 rays per probe for a total ray budget of 800-1600k (depending on scene) rays per frame. 

\begin{table*}[t!]
\centering
\caption{\label{tab:cacheDetails} Table showing probe encoding details for the various techniques we use in our comparison.}
\vspace{-8pt}
\begin{tabular}{|c|c|c|c|}
\hline
Technique & \begin{tabular}[c]{@{}c@{}}Irradiance \\ Cache Encoding\end{tabular} & \begin{tabular}[c]{@{}c@{}}Visibility\\ Cache Encoding\end{tabular} & \begin{tabular}[c]{@{}c@{}}Temporal\\ Hysteresis\end{tabular} \\ \hline
Ours & $\lfloor \text{R9} \rfloor \lfloor \text{G9} \rfloor \lfloor \text{B8} \rfloor - \text{N}$ & $[\text{R13}] [\text{G13}] - \text{N}$ & \begin{tabular}[c]{@{}c@{}}Static: 0.98 ($N_{max}=63$)\\ Dyna: 0.91 ($N_{max}=10$)\end{tabular} \\ \hline
Q-DDGI & $\lfloor \text{R11} \rfloor \lfloor \text{G11} \rfloor \lfloor \text{B10} \rfloor - \text{N}$ & $[\text{R16}] [\text{G16}] - \text{N}$ & 0.94 \\ \hline
Reference & RGB32f & RG32f & N/A \\ \hline
\end{tabular}
\end{table*}

\textit{Reference:} Reference implementation uses a standard \textit{FP32} representation for irradiance and visibility caches as shown in table \ref{tab:cacheDetails}. We also use a higher resolution $32\times32$ irradiance and visibility cache. Due to memory constraints, we are limited to a smaller probe-grid of size $32\times32\times32$ using same probe spacing (table \ref{tab:probeDetails}) as other techniques. For each frame, we discard any previous values in the probes and accumulate samples using a true-average with 64 rays per texel.

\textit{Ours:} We use 4096 instances of static chain invocations and 1024 instances of dynamic chain invocations. Overall, we use use between 500-900k (depending on scene) rays per frame.

Figure \ref{fig:teaser} and \ref{fig:bistroExtClose} shows a large scene (\textsc{Bistro Exterior}), with the tunnel's entry and exit modified with dynamic gates. The tunnel interior walls are illuminated by indirect illumination alone, controlled by the direct light bouncing off the floor. The direct illumination on the floor is controlled by the dynamic entry gate. The scene tests the tracking capabilities of our algorithm; the dynamic Markov-chain should sample the probes close to the moving door. The scene also tests our color compression scheme under low-light and moving-average accumulation.

Figure \ref{fig:spnzaDynEmit} shows the \textsc{Sponza} scene under dynamic lighting, testing the detection capabilities of ADGI in the absence of dynamic geometry. Figure \ref{fig:bistroExtStatic} shows a static scene without dynamic geometry or lighting, testing the convergence of our static adaptive sampling when no dynamism is detected or the dynamic changes are too slow to detect, such as day-night cycles in games. 

Figure \ref{fig:spnzaDynBuddha} shows a dynamic geometry (\textsc{Stanford Buddha}) under glossy indirect illumination with ambient lighting as direct component. The scene is stressful as the camera frustum contains many times more probes compared to other scenes due to the increased probe density required for glossy illumination. This scene tests the transient response of a dynamic geometry on a glossy floor. Thus the scene is less forgiving of spatio-temporal blurring. 

We measured the results on a desktop with Nvidia 2080Ti GPU and AMD 5600X CPU at $1920\times1080$ resolution. The performance numbers cited in table \ref{tab:performanceDetails} are only for ADGI and Q-DDGI algorithms. The GBuffer and direct-illumination passes require an additional 2ms and 3ms, respectively.

\section{Limitations \label{sec:limitations}}

We inherit similar limitations as the vanilla DDGI algorithm. The probe visibility from a shade-point is only approximate and requires modifications such as probe movement to minimize light leakage. The probe representation is not efficient in capturing glossy light-transport and requires a dense spatio-angular discretization of irradiance cache to capture glossy reflections.

Accurate detection of transient spatio-temporal changes in a scene are difficult. The accuracy of detecting dynamic geometry reduces with the distance of the dynamic object from a probe. The same is true for dynamic lighting; especially high frequency localized lighting that is far from a probe is difficult to detect. Also, for the Markov chain to track the target distribution, the speed of motion should be capped comparable to the product of Markov-chain step size and average frame-rate. \textcolor{black}{While many game engines keep track of the dynamic objects, facilitating the detection of changing in visibility, we still need ray-tracing to detect dynamic radiosity.} 

\begin{table}[t!]
\centering
\caption{\label{tab:performanceDetails} Performance breakdown of our technique and Q-DDGI. Our probe sampling stage is divided into three sub-stages - heuristic construction (P), static adaptive sampling (S), and dynamic adaptive sampling (D).}
\vspace{-8pt}
\begin{tabular}{|c|ccc|ccc|}
\hline
\multirow{2}{*}{Scene} & \multicolumn{3}{c|}{Ours (in milliseconds)} & \multicolumn{3}{c|}{Q-DDGI (in milliseconds)} \\ \cline{2-7} 
 & \multicolumn{1}{c|}{\begin{tabular}[c]{@{}c@{}}Probe Sampling\\ (P + S + D)\end{tabular}} & \multicolumn{1}{c|}{Deferred} & Total & \multicolumn{1}{c|}{Probe sampling} & \multicolumn{1}{c|}{Deferred} & Total \\ \hline
Bistro - Exterior & \multicolumn{1}{c|}{\begin{tabular}[c]{@{}c@{}}4.01 + 2.23 + 4.73\\ = 11.0\end{tabular}} & \multicolumn{1}{c|}{4.63} & 15.6 & \multicolumn{1}{c|}{22.3} & \multicolumn{1}{c|}{4.47} & 26.8 \\ \hline
Sponza - Diffuse & \multicolumn{1}{c|}{\begin{tabular}[c]{@{}c@{}}1.21 + 1.85 + 3.18\\ =6.24\end{tabular}} & \multicolumn{1}{c|}{3.62} & 9.86 & \multicolumn{1}{c|}{9.69} & \multicolumn{1}{c|}{3.51} & 13.2 \\ \hline
Sponza - Glossy & \multicolumn{1}{c|}{\begin{tabular}[c]{@{}c@{}}4.83 + 2.11 + 4.33\\ =11.27\end{tabular}} & \multicolumn{1}{c|}{6.44} & 19.7 & \multicolumn{1}{c|}{24.9} & \multicolumn{1}{c|}{6.71} & 31.6 \\ \hline
\end{tabular}
\end{table}

\section{Conclusion \label{sec:conclusion}}

Our adaptive sampling approach improves upon the efficiency of the original DDGI algorithm. Our approach non-uniformly allocates resources in regions with time varying phenomena and captures transient localized changes in an environment containing millions of probes. By contrast, DDGI's uniform allocation policy dilutes resource concentration in critical regions, especially when a large number of probes are present. These improvements reduce temporal lag and minimizes reliance on temporal blur to reduce noise. Our probe encoding scheme minimizes memory requirements by 4x (and by extension memory bandwidth) with minimal impact on quality while also enabling millions of probes in a scene. Our adaptive sampling stages have a fixed upper bound on the compute requirement and also decouples sampling from the number of probes, further reducing memory bandwidth requirement. These changes enable improved probe-based rendering while also enabling 1.5-2x performance improvements.

\section{Related work extension \label{sect:relatedWorkExtension}}

\textbf{Irradiance caching} 
Irradiance caching is another line of techniques attempting to overcome the high computation cost of GI. The irradiance caching method assumes that irradiance vary smoothly across the scene, and texture detail can be recovered using albedo modulation \citep{RadianceOld}. The interpolation and location of the various cache records is a critical, especially when the assumptions on smoothness do not hold. While robust, principled offline solutions exist \citep{RadianceMedia, RadiancePractical}, real-time applications often resort to complex heuristics and impose harsh constraints to achieve online GI. Compression \citep{RadianceCompress}, sparse interpolation \citep{RadianceSparse}, pre-convolved environment maps \citep{RadianceCluster, RadianceConvolve}, spatial hashing \citep{RadiancePath} and using neural network \citep{RadianceNeural} are instances of advancements in real-time irradiance caching. Although these approaches aim for real-time performance, their complexity and constraints make them challenging to implement and deploy.

\textbf{Path tracing}
The flexibility and generality offered by path tracing \citep{Rendering} is highly desirable for real-time rendering. However, path tracing has been out of reach for real-time applications due to its substantial computational requirements. Even with the advent of hardware-accelerated ray tracing \citep{HWRT09}, it is only possible to trace a few tens of rays at each pixel in real-time. Therefore, effective sampling strategies and high-quality denoising algorithms \citep{SVGF17, ASVGF2019, munkberg2020neural} are essential. Many sampling methods try to learn the representation of incident illumination during rendering \citep{SamplingVorba, Sampling2, Sampling3, Sampling4, Sampling5}. While these approaches can provide substantial error reduction, constructing these structures in parallel on a GPU incurs a significant overhead that seem unsuitable for real-time applications. Recently proposed ReSTIR GI \citep{RestirGI} provides an efficient real-time sampling strategy by reusing the paths spatially and temporally but the algorithm becomes complicated after second bounce and still requires denoising for the final stage. Deep learning has also been applied to path guiding, including work by \citep{NRC2020, NeuralSampling}. These approaches demonstrated a substantial reduction in error due to more effective path sampling, though their performance remain insufficient for real-time applications. 

\textbf{Screen space approaches:}
Approximating physically plausible illumination at real-time frame rates with screen space methods is popular in games. Screen space methods are fast, GPU-friendly, and simple to implement. Screen space ambient occlusion (SSAO) \citep{mittring2007finding, bavoil2008image} is part of many real-time rendering engines. Following SSAO, screen Space Directional Occlusion (SSDO) \citep{ritschel2009approximating} is used for near-field direct and indirect diffuse lighting. Sousa et al. \citep{sousa2011secrets} proposed Screen Space Reflections (SSR) using a 2D ray-tracing approach directly in screen space to obtain the indirect specular component. Recently Screen-Space Global Illumination (SSGI) \citep{silvennoinen2015multi,sousa2011secrets, ritschel2009approximating} methods offer a viable solution to real-time GI. However, these methods are limited by the information visible from the observer's position, thus making it difficult to engineer a robust solution.

\textbf{Importance sampling and Bayesian modeling:} Importance sampling provides a tool to reduce the cost of brute force integration by selectively evaluating elements of the integrand based on prior knowledge, i.e. an educated guess. Previous works in importance sampling proposed different methods to apply importance sampling to various Monte-Carlo integration existing in rendering equations \citep{sen2012filtering, keller2015path, veach1995optimally}. Although  Markov Chain Monte Carlo(MCMC) methods have been used in Bayesian learning from the early days of neural networks \citep{neal2012bayesian}, and Stochastic-Gradient MCMC has been proposed \citep{welling2011bayesian} with various applications \citep{li2016learning}, our approach is neither Monte Carlo-based nor Neural-network learning. We exploit Bayesian inference and Markov Chains as our mathematical means to sample the important texels on the probe, by defining our guide function (prior), likelihood, and posterior.

\textbf{Markov Chain:} 
Markov Chains are used broadly in Monte Carlo path-tracing. For example, Veach and Guibas \citep{veach1997metropolis} used Metropolis Sampling to explore the space of all possible paths. Kelemen et al. \citep{kelemen2002simple} later applied the exact sampling in the space of random numbers, i.e., in Primary Sample Space. The most recent work by Bitterli et al. \citep{bitterli2019selectively} combines a simple path tracing integrator with MCMC by using the random seeds of high variance paths as starting points for the Markov Chains. Although Markov Chains are encountered extensively beneficial in solving Monte Carlo sampling, our point of view on sampling and employing the Markov Chain to draw samples from the guide function is distinct.
 
\textbf{Bayesian inference:} 
Bayesian modeling is a widespread methodology in computer vision and graphics. Brouillat et al. \citep{brouillat2009bayesian} and Marques et al. \citep{marques2013spherical} pioneered the use of Bayesian Monte Carlo (BMC) \citep{rasmussen2003bayesian} in light transport simulation. In contrast, \citep{vevoda2018bayesian} keep the efficient classic, frequentist MC approach and apply Bayesian modeling to optimize their sampling distributions for direct illumination estimates across the scene. Similar approach is used by Vorba et al. \citep{vorba2014line}, who employ a maximum a posteriori (MAP) formulation to regularize training of parametric mixture models for optimized indirect illumination sampling. Our approach uses Bayesian modeling in the context of light-probes to detect important probes and directions based on sampled evidence.
\bibliographystyle{ACM-Reference-Format}
\bibliography{main}


\begin{thebibliography}{66}


\ifx \showCODEN    \undefined \def \showCODEN     #1{\unskip}     \fi
\ifx \showDOI      \undefined \def \showDOI       #1{#1}\fi
\ifx \showISBNx    \undefined \def \showISBNx     #1{\unskip}     \fi
\ifx \showISBNxiii \undefined \def \showISBNxiii  #1{\unskip}     \fi
\ifx \showISSN     \undefined \def \showISSN      #1{\unskip}     \fi
\ifx \showLCCN     \undefined \def \showLCCN      #1{\unskip}     \fi
\ifx \shownote     \undefined \def \shownote      #1{#1}          \fi
\ifx \showarticletitle \undefined \def \showarticletitle #1{#1}   \fi
\ifx \showURL      \undefined \def \showURL       {\relax}        \fi
\providecommand\bibfield[2]{#2}
\providecommand\bibinfo[2]{#2}
\providecommand\natexlab[1]{#1}
\providecommand\showeprint[2][]{arXiv:#2}

\bibitem[Sam(2019)]%
        {Sampling3}
 \bibinfo{year}{2019}\natexlab{}.
\newblock \bibinfo{booktitle}{\emph{SIGGRAPH '19: ACM SIGGRAPH 2019 Production
  Sessions}} (Los Angeles, California). \bibinfo{publisher}{Association for
  Computing Machinery}, \bibinfo{address}{New York, NY, USA}.
\newblock
\showISBNx{9781450361958}


\bibitem[Bavoil et~al\mbox{.}(2008)]%
        {bavoil2008image}
\bibfield{author}{\bibinfo{person}{Louis Bavoil}, \bibinfo{person}{Miguel
  Sainz}, {and} \bibinfo{person}{Rouslan Dimitrov}.}
  \bibinfo{year}{2008}\natexlab{}.
\newblock \showarticletitle{Image-Space Horizon-Based Ambient Occlusion}. In
  \bibinfo{booktitle}{\emph{ACM SIGGRAPH 2008 Talks}} (Los Angeles, California)
  \emph{(\bibinfo{series}{SIGGRAPH '08})}. \bibinfo{publisher}{Association for
  Computing Machinery}, \bibinfo{address}{New York, NY, USA}, Article
  \bibinfo{articleno}{22}, \bibinfo{numpages}{1}~pages.
\newblock
\showISBNx{9781605583433}
\urldef\tempurl%
\url{https://doi.org/10.1145/1401032.1401061}
\showDOI{\tempurl}


\bibitem[Binder et~al\mbox{.}(2018)]%
        {RadiancePath}
\bibfield{author}{\bibinfo{person}{Nikolaus Binder}, \bibinfo{person}{Sascha
  Fricke}, {and} \bibinfo{person}{Alexander Keller}.}
  \bibinfo{year}{2018}\natexlab{}.
\newblock \showarticletitle{Fast Path Space Filtering by Jittered Spatial
  Hashing}. In \bibinfo{booktitle}{\emph{ACM SIGGRAPH 2018 Talks}} (Vancouver,
  British Columbia, Canada) \emph{(\bibinfo{series}{SIGGRAPH '18})}.
  \bibinfo{publisher}{Association for Computing Machinery},
  \bibinfo{address}{New York, NY, USA}, Article \bibinfo{articleno}{71},
  \bibinfo{numpages}{2}~pages.
\newblock
\showISBNx{9781450358200}
\urldef\tempurl%
\url{https://doi.org/10.1145/3214745.3214806}
\showDOI{\tempurl}


\bibitem[Bitterli and Jarosz(2019)]%
        {bitterli2019selectively}
\bibfield{author}{\bibinfo{person}{Benedikt Bitterli} {and}
  \bibinfo{person}{Wojciech Jarosz}.} \bibinfo{year}{2019}\natexlab{}.
\newblock \showarticletitle{Selectively metropolised Monte Carlo light
  transport simulation}.
\newblock \bibinfo{journal}{\emph{ACM Transactions on Graphics (TOG)}}
  \bibinfo{volume}{38}, \bibinfo{number}{6} (\bibinfo{year}{2019}),
  \bibinfo{pages}{1--10}.
\newblock


\bibitem[Brouillat et~al\mbox{.}(2009)]%
        {brouillat2009bayesian}
\bibfield{author}{\bibinfo{person}{Jonathan Brouillat},
  \bibinfo{person}{Christian Bouville}, \bibinfo{person}{Brad Loos},
  \bibinfo{person}{Charles Hansen}, {and} \bibinfo{person}{Kadi Bouatouch}.}
  \bibinfo{year}{2009}\natexlab{}.
\newblock \showarticletitle{A Bayesian Monte Carlo approach to global
  illumination}. In \bibinfo{booktitle}{\emph{Computer Graphics Forum}},
  Vol.~\bibinfo{volume}{28}. \bibinfo{publisher}{Wiley Online Library},
  \bibinfo{address}{USA}, \bibinfo{pages}{2315--2329}.
\newblock


\bibitem[Chaitanya et~al\mbox{.}(2017)]%
        {ChaitanyaDenoise17}
\bibfield{author}{\bibinfo{person}{Chakravarty R.~Alla Chaitanya},
  \bibinfo{person}{Anton~S. Kaplanyan}, \bibinfo{person}{Christoph Schied},
  \bibinfo{person}{Marco Salvi}, \bibinfo{person}{Aaron Lefohn},
  \bibinfo{person}{Derek Nowrouzezahrai}, {and} \bibinfo{person}{Timo Aila}.}
  \bibinfo{year}{2017}\natexlab{}.
\newblock \showarticletitle{Interactive Reconstruction of Monte Carlo Image
  Sequences Using a Recurrent Denoising Autoencoder}.
\newblock \bibinfo{journal}{\emph{ACM Trans. Graph.}} \bibinfo{volume}{36},
  \bibinfo{number}{4}, Article \bibinfo{articleno}{98} (\bibinfo{date}{July}
  \bibinfo{year}{2017}), \bibinfo{numpages}{12}~pages.
\newblock
\showISSN{0730-0301}
\urldef\tempurl%
\url{https://doi.org/10.1145/3072959.3073601}
\showDOI{\tempurl}


\bibitem[Chib and Greenberg(1995)]%
        {chib1995understanding}
\bibfield{author}{\bibinfo{person}{Siddhartha Chib} {and}
  \bibinfo{person}{Edward Greenberg}.} \bibinfo{year}{1995}\natexlab{}.
\newblock \showarticletitle{Understanding the metropolis-hastings algorithm}.
\newblock \bibinfo{journal}{\emph{The american statistician}}
  \bibinfo{volume}{49}, \bibinfo{number}{4} (\bibinfo{year}{1995}),
  \bibinfo{pages}{327--335}.
\newblock


\bibitem[Diolatzis et~al\mbox{.}(2020)]%
        {Sampling4}
\bibfield{author}{\bibinfo{person}{Stavros Diolatzis}, \bibinfo{person}{Adrien
  Gruson}, \bibinfo{person}{Wenzel Jakob}, \bibinfo{person}{Derek
  Nowrouzezahrai}, {and} \bibinfo{person}{George Drettakis}.}
  \bibinfo{year}{2020}\natexlab{}.
\newblock \showarticletitle{Practical Product Path Guiding Using Linearly
  Transformed Cosines}.
\newblock \bibinfo{journal}{\emph{In Computer Graphics Forum (Proceedings of
  Eurographics Symposium on Rendering)}} \bibinfo{volume}{39},
  \bibinfo{number}{4} (\bibinfo{date}{July} \bibinfo{year}{2020}).
\newblock


\bibitem[Donnelly and Lauritzen(2006)]%
        {VSM06}
\bibfield{author}{\bibinfo{person}{William Donnelly} {and}
  \bibinfo{person}{Andrew Lauritzen}.} \bibinfo{year}{2006}\natexlab{}.
\newblock \showarticletitle{Variance Shadow Maps}. In
  \bibinfo{booktitle}{\emph{Proceedings of the 2006 Symposium on Interactive 3D
  Graphics and Games}} (Redwood City, California) \emph{(\bibinfo{series}{I3D
  '06})}. \bibinfo{publisher}{Association for Computing Machinery},
  \bibinfo{address}{New York, NY, USA}, \bibinfo{pages}{161–165}.
\newblock
\showISBNx{159593295X}
\urldef\tempurl%
\url{https://doi.org/10.1145/1111411.1111440}
\showDOI{\tempurl}


\bibitem[Geyer(1992)]%
        {geyer1992practical}
\bibfield{author}{\bibinfo{person}{Charles~J Geyer}.}
  \bibinfo{year}{1992}\natexlab{}.
\newblock \bibinfo{title}{Practical markov chain monte carlo}.
\newblock , \bibinfo{numpages}{473--483}~pages.
\newblock


\bibitem[Ghahramani and Rasmussen(2002)]%
        {rasmussen2003bayesian}
\bibfield{author}{\bibinfo{person}{Zoubin Ghahramani} {and}
  \bibinfo{person}{Carl Rasmussen}.} \bibinfo{year}{2002}\natexlab{}.
\newblock \bibinfo{title}{Bayesian Monte Carlo}.
\newblock
\newblock
\urldef\tempurl%
\url{https://proceedings.neurips.cc/paper/2002/file/24917db15c4e37e421866448c9ab23d8-Paper.pdf}
\showURL{%
\tempurl}


\bibitem[Hachisuka et~al\mbox{.}(2008)]%
        {toshiyaAdaptive08}
\bibfield{author}{\bibinfo{person}{Toshiya Hachisuka},
  \bibinfo{person}{Wojciech Jarosz}, \bibinfo{person}{Richard~Peter
  Weistroffer}, \bibinfo{person}{Kevin Dale}, \bibinfo{person}{Greg Humphreys},
  \bibinfo{person}{Matthias Zwicker}, {and} \bibinfo{person}{Henrik~Wann
  Jensen}.} \bibinfo{year}{2008}\natexlab{}.
\newblock \showarticletitle{Multidimensional Adaptive Sampling and
  Reconstruction for Ray Tracing}.
\newblock \bibinfo{journal}{\emph{ACM Trans. Graph.}} \bibinfo{volume}{27},
  \bibinfo{number}{3} (\bibinfo{date}{aug} \bibinfo{year}{2008}),
  \bibinfo{pages}{1–10}.
\newblock
\showISSN{0730-0301}
\urldef\tempurl%
\url{https://doi.org/10.1145/1360612.1360632}
\showDOI{\tempurl}


\bibitem[Hasselgren et~al\mbox{.}(2020)]%
        {neuralAdaptive20}
\bibfield{author}{\bibinfo{person}{Jon Hasselgren}, \bibinfo{person}{J.
  Munkberg}, \bibinfo{person}{Marco Salvi}, \bibinfo{person}{A. Patney}, {and}
  \bibinfo{person}{Aaron Lefohn}.} \bibinfo{year}{2020}\natexlab{}.
\newblock \showarticletitle{Neural Temporal Adaptive Sampling and Denoising}.
\newblock \bibinfo{journal}{\emph{Computer Graphics Forum}}
  \bibinfo{volume}{39} (\bibinfo{date}{05} \bibinfo{year}{2020}),
  \bibinfo{pages}{147--155}.
\newblock
\urldef\tempurl%
\url{https://doi.org/10.1111/cgf.13919}
\showDOI{\tempurl}


\bibitem[Ikkala et~al\mbox{.}(2021)]%
        {DDISH-GI}
\bibfield{author}{\bibinfo{person}{Julius Ikkala}, \bibinfo{person}{Petrus
  Kivi}, \bibinfo{person}{Joel Alanko}, \bibinfo{person}{Markku M\"{a}kitalo},
  {and} \bibinfo{person}{Pekka J\"{a}\"{a}skel\"{a}inen}.}
  \bibinfo{year}{2021}\natexlab{}.
\newblock \showarticletitle{DDISH-GI: Dynamic Distributed Spherical Harmonics
  Global Illumination}. In \bibinfo{booktitle}{\emph{Advances in Computer
  Graphics: 38th Computer Graphics International Conference, CGI 2021, Virtual
  Event, September 6–10, 2021, Proceedings}}.
  \bibinfo{publisher}{Springer-Verlag}, \bibinfo{address}{Berlin, Heidelberg},
  \bibinfo{pages}{433–451}.
\newblock
\showISBNx{978-3-030-89028-5}
\urldef\tempurl%
\url{https://doi.org/10.1007/978-3-030-89029-2_34}
\showDOI{\tempurl}


\bibitem[Iwanicki and Sloan(2017)]%
        {iwanicki2017precomputed}
\bibfield{author}{\bibinfo{person}{Micha{\l} Iwanicki} {and}
  \bibinfo{person}{Peter-Pike Sloan}.} \bibinfo{year}{2017}\natexlab{}.
\newblock \bibinfo{title}{Precomputed lighting in Call of Duty: Infinite
  Warfare}.
\newblock
\newblock


\bibitem[Jarosz et~al\mbox{.}(2008)]%
        {RadianceMedia}
\bibfield{author}{\bibinfo{person}{Wojciech Jarosz}, \bibinfo{person}{Craig
  Donner}, \bibinfo{person}{Matthias Zwicker}, {and}
  \bibinfo{person}{Henrik~Wann Jensen}.} \bibinfo{year}{2008}\natexlab{}.
\newblock \showarticletitle{Radiance Caching for Participating Media}.
\newblock \bibinfo{journal}{\emph{ACM Trans. Graph.}} \bibinfo{volume}{27},
  \bibinfo{number}{1}, Article \bibinfo{articleno}{7} (\bibinfo{date}{mar}
  \bibinfo{year}{2008}), \bibinfo{numpages}{11}~pages.
\newblock
\showISSN{0730-0301}
\urldef\tempurl%
\url{https://doi.org/10.1145/1330511.1330518}
\showDOI{\tempurl}


\bibitem[Jensen(1996)]%
        {jensen1996global}
\bibfield{author}{\bibinfo{person}{Henrik~Wann Jensen}.}
  \bibinfo{year}{1996}\natexlab{}.
\newblock \bibinfo{title}{Global illumination using photon maps}.
\newblock , \bibinfo{numpages}{21--30}~pages.
\newblock


\bibitem[Kajiya(1986)]%
        {Rendering}
\bibfield{author}{\bibinfo{person}{James~T. Kajiya}.}
  \bibinfo{year}{1986}\natexlab{}.
\newblock \showarticletitle{The Rendering Equation}. In
  \bibinfo{booktitle}{\emph{Proceedings of the 13th Annual Conference on
  Computer Graphics and Interactive Techniques}}
  \emph{(\bibinfo{series}{SIGGRAPH '86})}. \bibinfo{publisher}{Association for
  Computing Machinery}, \bibinfo{address}{New York, NY, USA},
  \bibinfo{pages}{143–150}.
\newblock
\showISBNx{0897911962}
\urldef\tempurl%
\url{https://doi.org/10.1145/15922.15902}
\showDOI{\tempurl}


\bibitem[Kelemen et~al\mbox{.}(2002)]%
        {kelemen2002simple}
\bibfield{author}{\bibinfo{person}{Csaba Kelemen},
  \bibinfo{person}{L{\'a}szl{\'o} Szirmay-Kalos}, \bibinfo{person}{Gy{\"o}rgy
  Antal}, {and} \bibinfo{person}{Ferenc Csonka}.}
  \bibinfo{year}{2002}\natexlab{}.
\newblock \bibinfo{title}{A simple and robust mutation strategy for the
  metropolis light transport algorithm}.
\newblock , \bibinfo{numpages}{531--540}~pages.
\newblock


\bibitem[Keller(1997)]%
        {keller1997instant}
\bibfield{author}{\bibinfo{person}{Alexander Keller}.}
  \bibinfo{year}{1997}\natexlab{}.
\newblock \showarticletitle{Instant Radiosity}. In
  \bibinfo{booktitle}{\emph{Proceedings of the 24th Annual Conference on
  Computer Graphics and Interactive Techniques}}
  \emph{(\bibinfo{series}{SIGGRAPH '97})}. \bibinfo{publisher}{ACM
  Press/Addison-Wesley Publishing Co.}, \bibinfo{address}{USA},
  \bibinfo{pages}{49–56}.
\newblock
\showISBNx{0897918967}
\urldef\tempurl%
\url{https://doi.org/10.1145/258734.258769}
\showDOI{\tempurl}


\bibitem[Keller et~al\mbox{.}(2015)]%
        {keller2015path}
\bibfield{author}{\bibinfo{person}{A. Keller}, \bibinfo{person}{L. Fascione},
  \bibinfo{person}{M. Fajardo}, \bibinfo{person}{I. Georgiev},
  \bibinfo{person}{P. Christensen}, \bibinfo{person}{J. Hanika},
  \bibinfo{person}{C. Eisenacher}, {and} \bibinfo{person}{G. Nichols}.}
  \bibinfo{year}{2015}\natexlab{}.
\newblock \showarticletitle{The Path Tracing Revolution in the Movie Industry}.
  In \bibinfo{booktitle}{\emph{ACM SIGGRAPH 2015 Courses}} (Los Angeles,
  California) \emph{(\bibinfo{series}{SIGGRAPH '15})}.
  \bibinfo{publisher}{Association for Computing Machinery},
  \bibinfo{address}{New York, NY, USA}, Article \bibinfo{articleno}{24},
  \bibinfo{numpages}{7}~pages.
\newblock
\showISBNx{9781450336345}
\urldef\tempurl%
\url{https://doi.org/10.1145/2776880.2792699}
\showDOI{\tempurl}


\bibitem[Keller et~al\mbox{.}(2019)]%
        {keller2019we}
\bibfield{author}{\bibinfo{person}{Alexander Keller}, \bibinfo{person}{Timo
  Viitanen}, \bibinfo{person}{Colin Barr\'{e}-Brisebois},
  \bibinfo{person}{Christoph Schied}, {and} \bibinfo{person}{Morgan McGuire}.}
  \bibinfo{year}{2019}\natexlab{}.
\newblock \showarticletitle{Are We Done with Ray Tracing?}. In
  \bibinfo{booktitle}{\emph{ACM SIGGRAPH 2019 Courses}} (Los Angeles,
  California) \emph{(\bibinfo{series}{SIGGRAPH '19})}.
  \bibinfo{publisher}{Association for Computing Machinery},
  \bibinfo{address}{New York, NY, USA}, Article \bibinfo{articleno}{3},
  \bibinfo{numpages}{381}~pages.
\newblock
\showISBNx{9781450363075}
\urldef\tempurl%
\url{https://doi.org/10.1145/3305366.3329896}
\showDOI{\tempurl}


\bibitem[Keller and Waechter(2009)]%
        {HWRT09}
\bibfield{author}{\bibinfo{person}{Alexander Keller} {and}
  \bibinfo{person}{Carsten Waechter}.} \bibinfo{year}{2009}\natexlab{}.
\newblock \bibinfo{title}{Real-time precision ray tracing}.
\newblock
\newblock
\urldef\tempurl%
\url{https://patents.google.com/patent/US20070024615A1/en}
\showURL{%
\tempurl}
\newblock
\shownote{US patent US20070024615A1}.


\bibitem[K\v{r}iv\'{a}nek et~al\mbox{.}(2008)]%
        {RadiancePractical}
\bibfield{author}{\bibinfo{person}{Jaroslav K\v{r}iv\'{a}nek},
  \bibinfo{person}{Pascal Gautron}, \bibinfo{person}{Greg Ward},
  \bibinfo{person}{Henrik~Wann Jensen}, \bibinfo{person}{Per~H. Christensen},
  {and} \bibinfo{person}{Eric Tabellion}.} \bibinfo{year}{2008}\natexlab{}.
\newblock \showarticletitle{Practical Global Illumination with Irradiance
  Caching}. In \bibinfo{booktitle}{\emph{ACM SIGGRAPH 2008 Classes}} (Los
  Angeles, California) \emph{(\bibinfo{series}{SIGGRAPH '08})}.
  \bibinfo{publisher}{Association for Computing Machinery},
  \bibinfo{address}{New York, NY, USA}, Article \bibinfo{articleno}{60},
  \bibinfo{numpages}{20}~pages.
\newblock
\showISBNx{9781450378451}
\urldef\tempurl%
\url{https://doi.org/10.1145/1401132.1401213}
\showDOI{\tempurl}


\bibitem[Li et~al\mbox{.}(2016)]%
        {li2016learning}
\bibfield{author}{\bibinfo{person}{Chunyuan Li}, \bibinfo{person}{Andrew
  Stevens}, \bibinfo{person}{Changyou Chen}, \bibinfo{person}{Yunchen Pu},
  \bibinfo{person}{Zhe Gan}, {and} \bibinfo{person}{Lawrence Carin}.}
  \bibinfo{year}{2016}\natexlab{}.
\newblock \bibinfo{title}{Learning Weight Uncertainty with Stochastic Gradient
  MCMC for Shape Classification}.
\newblock , \bibinfo{numpages}{5666-5675}~pages.
\newblock
\showISSN{1063-6919}
\urldef\tempurl%
\url{https://doi.org/10.1109/CVPR.2016.611}
\showDOI{\tempurl}


\bibitem[Lin et~al\mbox{.}(2021)]%
        {VolRestir2021}
\bibfield{author}{\bibinfo{person}{Daqi Lin}, \bibinfo{person}{Chris Wyman},
  {and} \bibinfo{person}{Cem Yuksel}.} \bibinfo{year}{2021}\natexlab{}.
\newblock \showarticletitle{Fast Volume Rendering with Spatiotemporal Reservoir
  Resampling}.
\newblock \bibinfo{journal}{\emph{ACM Trans. Graph.}} \bibinfo{volume}{40},
  \bibinfo{number}{6}, Article \bibinfo{articleno}{279} (\bibinfo{date}{dec}
  \bibinfo{year}{2021}), \bibinfo{numpages}{18}~pages.
\newblock
\showISSN{0730-0301}
\urldef\tempurl%
\url{https://doi.org/10.1145/3478513.3480499}
\showDOI{\tempurl}


\bibitem[Lin and Yuksel(2020)]%
        {StoLCuts2020}
\bibfield{author}{\bibinfo{person}{Daqi Lin} {and} \bibinfo{person}{Cem
  Yuksel}.} \bibinfo{year}{2020}\natexlab{}.
\newblock \showarticletitle{Real-Time Stochastic Lightcuts}.
\newblock \bibinfo{journal}{\emph{Proc. ACM Comput. Graph. Interact. Tech.}}
  \bibinfo{volume}{3}, \bibinfo{number}{1}, Article \bibinfo{articleno}{5}
  (\bibinfo{date}{apr} \bibinfo{year}{2020}), \bibinfo{numpages}{18}~pages.
\newblock
\urldef\tempurl%
\url{https://doi.org/10.1145/3384543}
\showDOI{\tempurl}


\bibitem[Majercik et~al\mbox{.}(2019)]%
        {DDGI19}
\bibfield{author}{\bibinfo{person}{Zander Majercik},
  \bibinfo{person}{Jean-Philippe Guertin}, \bibinfo{person}{Derek
  Nowrouzezahrai}, {and} \bibinfo{person}{Morgan McGuire}.}
  \bibinfo{year}{2019}\natexlab{}.
\newblock \showarticletitle{Dynamic Diffuse Global Illumination with Ray-Traced
  Irradiance Fields}.
\newblock \bibinfo{journal}{\emph{Journal of Computer Graphics Techniques
  (JCGT)}} \bibinfo{volume}{8}, \bibinfo{number}{2} (\bibinfo{date}{5 June}
  \bibinfo{year}{2019}), \bibinfo{pages}{1--30}.
\newblock
\showISSN{2331-7418}
\urldef\tempurl%
\url{http://jcgt.org/published/0008/02/01/}
\showURL{%
\tempurl}


\bibitem[Majercik et~al\mbox{.}(2021)]%
        {majercik2020scaling}
\bibfield{author}{\bibinfo{person}{Zander Majercik}, \bibinfo{person}{Adam
  Marrs}, \bibinfo{person}{Josef Spjut}, {and} \bibinfo{person}{Morgan
  McGuire}.} \bibinfo{year}{2021}\natexlab{}.
\newblock \showarticletitle{Scaling Probe-Based Real-Time Dynamic Global
  Illumination for Production}.
\newblock \bibinfo{journal}{\emph{Journal of Computer Graphics Techniques
  (JCGT)}} \bibinfo{volume}{10}, \bibinfo{number}{2} (\bibinfo{date}{3 May}
  \bibinfo{year}{2021}), \bibinfo{pages}{1--29}.
\newblock
\showISSN{2331-7418}
\urldef\tempurl%
\url{http://jcgt.org/published/0010/02/01/}
\showURL{%
\tempurl}


\bibitem[Marques et~al\mbox{.}(2013)]%
        {marques2013spherical}
\bibfield{author}{\bibinfo{person}{Ricardo Marques}, \bibinfo{person}{Christian
  Bouville}, \bibinfo{person}{Mickaël Ribardière},
  \bibinfo{person}{Luís~Paulo Santos}, {and} \bibinfo{person}{Kadi
  Bouatouch}.} \bibinfo{year}{2013}\natexlab{}.
\newblock \showarticletitle{A Spherical Gaussian Framework for Bayesian Monte
  Carlo Rendering of Glossy Surfaces}.
\newblock \bibinfo{journal}{\emph{IEEE Transactions on Visualization and
  Computer Graphics}} \bibinfo{volume}{19}, \bibinfo{number}{10}
  (\bibinfo{year}{2013}), \bibinfo{pages}{1619--1632}.
\newblock
\urldef\tempurl%
\url{https://doi.org/10.1109/TVCG.2013.79}
\showDOI{\tempurl}


\bibitem[McGuire et~al\mbox{.}(2017)]%
        {mcguire2017real}
\bibfield{author}{\bibinfo{person}{Morgan McGuire}, \bibinfo{person}{Mike
  Mara}, \bibinfo{person}{Derek Nowrouzezahrai}, {and} \bibinfo{person}{David
  Luebke}.} \bibinfo{year}{2017}\natexlab{}.
\newblock \showarticletitle{Real-Time Global Illumination Using Precomputed
  Light Field Probes}. In \bibinfo{booktitle}{\emph{Proceedings of the 21st ACM
  SIGGRAPH Symposium on Interactive 3D Graphics and Games}} (San Francisco,
  California) \emph{(\bibinfo{series}{I3D '17})}.
  \bibinfo{publisher}{Association for Computing Machinery},
  \bibinfo{address}{New York, NY, USA}, Article \bibinfo{articleno}{2},
  \bibinfo{numpages}{11}~pages.
\newblock
\showISBNx{9781450348867}
\urldef\tempurl%
\url{https://doi.org/10.1145/3023368.3023378}
\showDOI{\tempurl}


\bibitem[Mehta et~al\mbox{.}(2012)]%
        {AAF12}
\bibfield{author}{\bibinfo{person}{Soham~Uday Mehta}, \bibinfo{person}{Brandon
  Wang}, {and} \bibinfo{person}{Ravi Ramamoorthi}.}
  \bibinfo{year}{2012}\natexlab{}.
\newblock \showarticletitle{Axis-Aligned Filtering for Interactive Sampled Soft
  Shadows}.
\newblock \bibinfo{journal}{\emph{ACM Trans. Graph.}} \bibinfo{volume}{31},
  \bibinfo{number}{6}, Article \bibinfo{articleno}{163} (\bibinfo{date}{Nov.}
  \bibinfo{year}{2012}), \bibinfo{numpages}{10}~pages.
\newblock
\showISSN{0730-0301}
\urldef\tempurl%
\url{https://doi.org/10.1145/2366145.2366182}
\showDOI{\tempurl}


\bibitem[Mittring(2007)]%
        {mittring2007finding}
\bibfield{author}{\bibinfo{person}{Martin Mittring}.}
  \bibinfo{year}{2007}\natexlab{}.
\newblock \showarticletitle{Finding next Gen: CryEngine 2}. In
  \bibinfo{booktitle}{\emph{ACM SIGGRAPH 2007 Courses}} (San Diego, California)
  \emph{(\bibinfo{series}{SIGGRAPH '07})}. \bibinfo{publisher}{Association for
  Computing Machinery}, \bibinfo{address}{New York, NY, USA},
  \bibinfo{pages}{97–121}.
\newblock
\showISBNx{9781450318235}
\urldef\tempurl%
\url{https://doi.org/10.1145/1281500.1281671}
\showDOI{\tempurl}


\bibitem[M\"uller et~al\mbox{.}(2017)]%
        {Sampling2}
\bibfield{author}{\bibinfo{person}{Thomas M\"uller}, \bibinfo{person}{Markus
  Gross}, {and} \bibinfo{person}{Jan Nov\'ak}.}
  \bibinfo{year}{2017}\natexlab{}.
\newblock \showarticletitle{Practical Path Guiding for Efficient
  Light-Transport Simulation}.
\newblock \bibinfo{journal}{\emph{Computer Graphics Forum (Proceedings of
  EGSR)}} \bibinfo{volume}{36}, \bibinfo{number}{4} (\bibinfo{date}{June}
  \bibinfo{year}{2017}), \bibinfo{pages}{91--100}.
\newblock
\urldef\tempurl%
\url{https://doi.org/10.1111/cgf.13227}
\showDOI{\tempurl}


\bibitem[M\"{u}ller et~al\mbox{.}(2019)]%
        {NeuralSampling}
\bibfield{author}{\bibinfo{person}{Thomas M\"{u}ller}, \bibinfo{person}{Brian
  Mcwilliams}, \bibinfo{person}{Fabrice Rousselle}, \bibinfo{person}{Markus
  Gross}, {and} \bibinfo{person}{Jan Nov\'{a}k}.}
  \bibinfo{year}{2019}\natexlab{}.
\newblock \showarticletitle{Neural Importance Sampling}.
\newblock \bibinfo{journal}{\emph{ACM Trans. Graph.}} \bibinfo{volume}{38},
  \bibinfo{number}{5}, Article \bibinfo{articleno}{145} (\bibinfo{date}{oct}
  \bibinfo{year}{2019}), \bibinfo{numpages}{19}~pages.
\newblock
\showISSN{0730-0301}
\urldef\tempurl%
\url{https://doi.org/10.1145/3341156}
\showDOI{\tempurl}


\bibitem[M\"{u}ller et~al\mbox{.}(2021a)]%
        {NRC2020}
\bibfield{author}{\bibinfo{person}{Thomas M\"{u}ller}, \bibinfo{person}{Fabrice
  Rousselle}, \bibinfo{person}{Jan Nov\'{a}k}, {and} \bibinfo{person}{Alexander
  Keller}.} \bibinfo{year}{2021}\natexlab{a}.
\newblock \showarticletitle{Real-Time Neural Radiance Caching for Path
  Tracing}.
\newblock \bibinfo{journal}{\emph{ACM Trans. Graph.}} \bibinfo{volume}{40},
  \bibinfo{number}{4}, Article \bibinfo{articleno}{36} (\bibinfo{date}{jul}
  \bibinfo{year}{2021}), \bibinfo{numpages}{16}~pages.
\newblock
\showISSN{0730-0301}
\urldef\tempurl%
\url{https://doi.org/10.1145/3450626.3459812}
\showDOI{\tempurl}


\bibitem[M\"{u}ller et~al\mbox{.}(2021b)]%
        {RadianceNeural}
\bibfield{author}{\bibinfo{person}{Thomas M\"{u}ller}, \bibinfo{person}{Fabrice
  Rousselle}, \bibinfo{person}{Jan Nov\'{a}k}, {and} \bibinfo{person}{Alexander
  Keller}.} \bibinfo{year}{2021}\natexlab{b}.
\newblock \showarticletitle{Real-Time Neural Radiance Caching for Path
  Tracing}.
\newblock \bibinfo{journal}{\emph{ACM Trans. Graph.}} \bibinfo{volume}{40},
  \bibinfo{number}{4}, Article \bibinfo{articleno}{36} (\bibinfo{date}{jul}
  \bibinfo{year}{2021}), \bibinfo{numpages}{16}~pages.
\newblock
\showISSN{0730-0301}
\urldef\tempurl%
\url{https://doi.org/10.1145/3450626.3459812}
\showDOI{\tempurl}


\bibitem[Munkberg and Hasselgren(2020)]%
        {munkberg2020neural}
\bibfield{author}{\bibinfo{person}{Jacob Munkberg} {and} \bibinfo{person}{Jon
  Hasselgren}.} \bibinfo{year}{2020}\natexlab{}.
\newblock \showarticletitle{Neural denoising with layer embeddings}. In
  \bibinfo{booktitle}{\emph{Computer Graphics Forum}},
  Vol.~\bibinfo{volume}{39}. Wiley Online Library, \bibinfo{pages}{1--12}.
\newblock


\bibitem[Neal(2012)]%
        {neal2012bayesian}
\bibfield{author}{\bibinfo{person}{Radford~M Neal}.}
  \bibinfo{year}{2012}\natexlab{}.
\newblock \bibinfo{booktitle}{\emph{Bayesian learning for neural networks}}.
  Vol.~\bibinfo{volume}{118}.
\newblock \bibinfo{publisher}{Springer Science \& Business Media},
  \bibinfo{address}{USA}.
\newblock


\bibitem[Ouyang et~al\mbox{.}(2021a)]%
        {RestirGI2021}
\bibfield{author}{\bibinfo{person}{Y. Ouyang}, \bibinfo{person}{S. Liu},
  \bibinfo{person}{M. Kettunen}, \bibinfo{person}{M. Pharr}, {and}
  \bibinfo{person}{J. Pantaleoni}.} \bibinfo{year}{2021}\natexlab{a}.
\newblock \showarticletitle{ReSTIR GI: Path Resampling for Real-Time Path
  Tracing}.
\newblock \bibinfo{journal}{\emph{Computer Graphics Forum}}
  \bibinfo{volume}{40}, \bibinfo{number}{8} (\bibinfo{year}{2021}),
  \bibinfo{pages}{17--29}.
\newblock
\urldef\tempurl%
\url{https://doi.org/10.1111/cgf.14378}
\showDOI{\tempurl}
\showeprint{https://onlinelibrary.wiley.com/doi/pdf/10.1111/cgf.14378}


\bibitem[Ouyang et~al\mbox{.}(2021b)]%
        {RestirGI}
\bibfield{author}{\bibinfo{person}{Yaobin Ouyang}, \bibinfo{person}{Shiqiu
  Liu}, \bibinfo{person}{Markus Kettunen}, \bibinfo{person}{Matt Pharr}, {and}
  \bibinfo{person}{Jacopo Pantaleoni}.} \bibinfo{year}{2021}\natexlab{b}.
\newblock \showarticletitle{ReSTIR GI: Path Resampling for Real-Time Path
  Tracing}. In \bibinfo{booktitle}{\emph{Computer Graphics Forum}},
  Vol.~\bibinfo{volume}{40}. Wiley Online Library, \bibinfo{pages}{17--29}.
\newblock


\bibitem[Rehfeld et~al\mbox{.}(2014)]%
        {RadianceCluster}
\bibfield{author}{\bibinfo{person}{Hauke Rehfeld}, \bibinfo{person}{Tobias
  Zirr}, {and} \bibinfo{person}{Carsten Dachsbacher}.}
  \bibinfo{year}{2014}\natexlab{}.
\newblock \showarticletitle{Clustered Pre-Convolved Radiance Caching}. In
  \bibinfo{booktitle}{\emph{Proceedings of the 14th Eurographics Symposium on
  Parallel Graphics and Visualization}} (Swansea, Wales, United Kingdom)
  \emph{(\bibinfo{series}{PGV '14})}. \bibinfo{publisher}{Eurographics
  Association}, \bibinfo{address}{Goslar, DEU}, \bibinfo{pages}{25–32}.
\newblock
\showISBNx{9783905674590}


\bibitem[Ritschel et~al\mbox{.}(2009)]%
        {ritschel2009approximating}
\bibfield{author}{\bibinfo{person}{Tobias Ritschel}, \bibinfo{person}{Thorsten
  Grosch}, {and} \bibinfo{person}{Hans-Peter Seidel}.}
  \bibinfo{year}{2009}\natexlab{}.
\newblock \showarticletitle{Approximating Dynamic Global Illumination in Image
  Space}. In \bibinfo{booktitle}{\emph{Proceedings of the 2009 Symposium on
  Interactive 3D Graphics and Games}} (Boston, Massachusetts)
  \emph{(\bibinfo{series}{I3D '09})}. \bibinfo{publisher}{Association for
  Computing Machinery}, \bibinfo{address}{New York, NY, USA},
  \bibinfo{pages}{75–82}.
\newblock
\showISBNx{9781605584294}
\urldef\tempurl%
\url{https://doi.org/10.1145/1507149.1507161}
\showDOI{\tempurl}


\bibitem[Ruppert et~al\mbox{.}(2020)]%
        {Sampling5}
\bibfield{author}{\bibinfo{person}{Lukas Ruppert}, \bibinfo{person}{Sebastian
  Herholz}, {and} \bibinfo{person}{Hendrik P.~A. Lensch}.}
  \bibinfo{year}{2020}\natexlab{}.
\newblock \showarticletitle{Robust Fitting of Parallax-Aware Mixtures for Path
  Guiding}.
\newblock \bibinfo{journal}{\emph{ACM Trans. Graph.}} \bibinfo{volume}{39},
  \bibinfo{number}{4}, Article \bibinfo{articleno}{147} (\bibinfo{date}{jul}
  \bibinfo{year}{2020}), \bibinfo{numpages}{15}~pages.
\newblock
\showISSN{0730-0301}
\urldef\tempurl%
\url{https://doi.org/10.1145/3386569.3392421}
\showDOI{\tempurl}


\bibitem[Scherzer et~al\mbox{.}(2012)]%
        {RadianceConvolve}
\bibfield{author}{\bibinfo{person}{Daniel Scherzer}, \bibinfo{person}{Chuong
  Nguyen}, \bibinfo{person}{Tobias Ritschel}, {and} \bibinfo{person}{Hans-Peter
  Seidel}.} \bibinfo{year}{2012}\natexlab{}.
\newblock \showarticletitle{{Pre-convolved Radiance Caching}}.
\newblock \bibinfo{journal}{\emph{Computer Graphics Forum}}
  (\bibinfo{year}{2012}).
\newblock
\showISSN{1467-8659}
\urldef\tempurl%
\url{https://doi.org/10.1111/j.1467-8659.2012.03134.x}
\showDOI{\tempurl}


\bibitem[Schied et~al\mbox{.}(2017)]%
        {SVGF17}
\bibfield{author}{\bibinfo{person}{Christoph Schied}, \bibinfo{person}{Anton
  Kaplanyan}, \bibinfo{person}{Chris Wyman}, \bibinfo{person}{Anjul Patney},
  \bibinfo{person}{Chakravarty R.~Alla Chaitanya}, \bibinfo{person}{John
  Burgess}, \bibinfo{person}{Shiqiu Liu}, \bibinfo{person}{Carsten
  Dachsbacher}, \bibinfo{person}{Aaron Lefohn}, {and} \bibinfo{person}{Marco
  Salvi}.} \bibinfo{year}{2017}\natexlab{}.
\newblock \showarticletitle{Spatiotemporal Variance-Guided Filtering: Real-Time
  Reconstruction for Path-Traced Global Illumination}. In
  \bibinfo{booktitle}{\emph{Proceedings of High Performance Graphics}} (Los
  Angeles, California) \emph{(\bibinfo{series}{HPG '17})}.
  \bibinfo{publisher}{Association for Computing Machinery},
  \bibinfo{address}{New York, NY, USA}, Article \bibinfo{articleno}{2},
  \bibinfo{numpages}{12}~pages.
\newblock
\showISBNx{9781450351010}
\urldef\tempurl%
\url{https://doi.org/10.1145/3105762.3105770}
\showDOI{\tempurl}


\bibitem[Schied et~al\mbox{.}(2018)]%
        {ASVGF2019}
\bibfield{author}{\bibinfo{person}{Christoph Schied},
  \bibinfo{person}{Christoph Peters}, {and} \bibinfo{person}{Carsten
  Dachsbacher}.} \bibinfo{year}{2018}\natexlab{}.
\newblock \showarticletitle{Gradient estimation for real-time adaptive temporal
  filtering}.
\newblock \bibinfo{journal}{\emph{Proceedings of the ACM on Computer Graphics
  and Interactive Techniques}} \bibinfo{volume}{1}, \bibinfo{number}{2}
  (\bibinfo{year}{2018}), \bibinfo{pages}{1--16}.
\newblock


\bibitem[Sen and Darabi(2012)]%
        {sen2012filtering}
\bibfield{author}{\bibinfo{person}{Pradeep Sen} {and} \bibinfo{person}{Soheil
  Darabi}.} \bibinfo{year}{2012}\natexlab{}.
\newblock \showarticletitle{On filtering the noise from the random parameters
  in Monte Carlo rendering.}
\newblock \bibinfo{journal}{\emph{ACM Trans. Graph.}} \bibinfo{volume}{31},
  \bibinfo{number}{3} (\bibinfo{year}{2012}), \bibinfo{pages}{18--1}.
\newblock


\bibitem[Silvennoinen and Lehtinen(2017)]%
        {RadianceSparse}
\bibfield{author}{\bibinfo{person}{Ari Silvennoinen} {and}
  \bibinfo{person}{Jaakko Lehtinen}.} \bibinfo{year}{2017}\natexlab{}.
\newblock \showarticletitle{Real-Time Global Illumination by Precomputed Local
  Reconstruction from Sparse Radiance Probes}.
\newblock \bibinfo{journal}{\emph{ACM Trans. Graph.}} \bibinfo{volume}{36},
  \bibinfo{number}{6}, Article \bibinfo{articleno}{230} (\bibinfo{date}{nov}
  \bibinfo{year}{2017}), \bibinfo{numpages}{13}~pages.
\newblock
\showISSN{0730-0301}
\urldef\tempurl%
\url{https://doi.org/10.1145/3130800.3130852}
\showDOI{\tempurl}


\bibitem[Silvennoinen and Timonen(2015)]%
        {silvennoinen2015multi}
\bibfield{author}{\bibinfo{person}{Ari Silvennoinen} {and}
  \bibinfo{person}{Ville Timonen}.} \bibinfo{year}{2015}\natexlab{}.
\newblock \bibinfo{title}{Multi-scale global illumination in quantum break}.
\newblock
\newblock


\bibitem[Sloan et~al\mbox{.}(2002)]%
        {sloan2002precomputed}
\bibfield{author}{\bibinfo{person}{Peter-Pike Sloan}, \bibinfo{person}{Jan
  Kautz}, {and} \bibinfo{person}{John Snyder}.}
  \bibinfo{year}{2002}\natexlab{}.
\newblock \showarticletitle{Precomputed Radiance Transfer for Real-Time
  Rendering in Dynamic, Low-Frequency Lighting Environments}. In
  \bibinfo{booktitle}{\emph{Proceedings of the 29th Annual Conference on
  Computer Graphics and Interactive Techniques}} (San Antonio, Texas)
  \emph{(\bibinfo{series}{SIGGRAPH '02})}. \bibinfo{publisher}{Association for
  Computing Machinery}, \bibinfo{address}{New York, NY, USA},
  \bibinfo{pages}{527–536}.
\newblock
\showISBNx{1581135211}
\urldef\tempurl%
\url{https://doi.org/10.1145/566570.566612}
\showDOI{\tempurl}


\bibitem[Sousa et~al\mbox{.}(2011)]%
        {sousa2011secrets}
\bibfield{author}{\bibinfo{person}{Tiago Sousa}, \bibinfo{person}{Nick Kasyan},
  {and} \bibinfo{person}{Nicolas Schulz}.} \bibinfo{year}{2011}\natexlab{}.
\newblock \bibinfo{title}{Secrets of CryENGINE 3 graphics technology}.
\newblock
\newblock


\bibitem[Stengel et~al\mbox{.}(2021)]%
        {stengel2021distributed}
\bibfield{author}{\bibinfo{person}{Michael Stengel}, \bibinfo{person}{Zander
  Majercik}, \bibinfo{person}{Benjamin Boudaoud}, {and} \bibinfo{person}{Morgan
  McGuire}.} \bibinfo{year}{2021}\natexlab{}.
\newblock \showarticletitle{A Distributed, Decoupled System for Losslessly
  Streaming Dynamic Light Probes to Thin Clients}. In
  \bibinfo{booktitle}{\emph{Proceedings of the 12th ACM Multimedia Systems
  Conference}} (Istanbul, Turkey) \emph{(\bibinfo{series}{MMSys '21})}.
  \bibinfo{publisher}{Association for Computing Machinery},
  \bibinfo{address}{New York, NY, USA}, \bibinfo{pages}{159–172}.
\newblock
\showISBNx{9781450384346}
\urldef\tempurl%
\url{https://doi.org/10.1145/3458305.3463379}
\showDOI{\tempurl}


\bibitem[Tabellion and Lamorlette(2004)]%
        {tabellion2004approximate}
\bibfield{author}{\bibinfo{person}{Eric Tabellion} {and}
  \bibinfo{person}{Arnauld Lamorlette}.} \bibinfo{year}{2004}\natexlab{}.
\newblock \showarticletitle{An Approximate Global Illumination System for
  Computer Generated Films}. In \bibinfo{booktitle}{\emph{ACM SIGGRAPH 2004
  Papers}} (Los Angeles, California) \emph{(\bibinfo{series}{SIGGRAPH '04})}.
  \bibinfo{publisher}{Association for Computing Machinery},
  \bibinfo{address}{New York, NY, USA}, \bibinfo{pages}{469–476}.
\newblock
\showISBNx{9781450378239}
\urldef\tempurl%
\url{https://doi.org/10.1145/1186562.1015748}
\showDOI{\tempurl}


\bibitem[Tatarchuk(2005)]%
        {tatarchuk2005irradiance}
\bibfield{author}{\bibinfo{person}{Natalya Tatarchuk}.}
  \bibinfo{year}{2005}\natexlab{}.
\newblock \bibinfo{title}{Irradiance volumes for games}.
\newblock , \bibinfo{numpages}{0}~pages.
\newblock


\bibitem[Vardis et~al\mbox{.}(2014)]%
        {RadianceCompress}
\bibfield{author}{\bibinfo{person}{Kostas Vardis}, \bibinfo{person}{Georgios
  Papaioannou}, {and} \bibinfo{person}{Anastasios Gkaravelis}.}
  \bibinfo{year}{2014}\natexlab{}.
\newblock \showarticletitle{Real-time Radiance Caching using Chrominance
  Compression}.
\newblock \bibinfo{journal}{\emph{Journal of Computer Graphics Techniques
  (JCGT)}} \bibinfo{volume}{3}, \bibinfo{number}{4} (\bibinfo{date}{16
  December} \bibinfo{year}{2014}), \bibinfo{pages}{111--131}.
\newblock
\showISSN{2331-7418}
\urldef\tempurl%
\url{http://jcgt.org/published/0003/04/06/}
\showURL{%
\tempurl}


\bibitem[Veach and Guibas(1995)]%
        {veach1995optimally}
\bibfield{author}{\bibinfo{person}{Eric Veach} {and}
  \bibinfo{person}{Leonidas~J. Guibas}.} \bibinfo{year}{1995}\natexlab{}.
\newblock \showarticletitle{Optimally Combining Sampling Techniques for Monte
  Carlo Rendering}. In \bibinfo{booktitle}{\emph{Proceedings of the 22nd Annual
  Conference on Computer Graphics and Interactive Techniques}}
  \emph{(\bibinfo{series}{SIGGRAPH '95})}. \bibinfo{publisher}{Association for
  Computing Machinery}, \bibinfo{address}{New York, NY, USA},
  \bibinfo{pages}{419–428}.
\newblock
\showISBNx{0897917014}
\urldef\tempurl%
\url{https://doi.org/10.1145/218380.218498}
\showDOI{\tempurl}


\bibitem[Veach and Guibas(1997)]%
        {veach1997metropolis}
\bibfield{author}{\bibinfo{person}{Eric Veach} {and}
  \bibinfo{person}{Leonidas~J. Guibas}.} \bibinfo{year}{1997}\natexlab{}.
\newblock \showarticletitle{Metropolis Light Transport}. In
  \bibinfo{booktitle}{\emph{Proceedings of the 24th Annual Conference on
  Computer Graphics and Interactive Techniques}}
  \emph{(\bibinfo{series}{SIGGRAPH '97})}. \bibinfo{publisher}{ACM
  Press/Addison-Wesley Publishing Co.}, \bibinfo{address}{USA},
  \bibinfo{pages}{65–76}.
\newblock
\showISBNx{0897918967}
\urldef\tempurl%
\url{https://doi.org/10.1145/258734.258775}
\showDOI{\tempurl}


\bibitem[V{\'e}voda et~al\mbox{.}(2018)]%
        {vevoda2018bayesian}
\bibfield{author}{\bibinfo{person}{Petr V{\'e}voda}, \bibinfo{person}{Ivo
  Kondapaneni}, {and} \bibinfo{person}{Jaroslav K{\v{r}}iv{\'a}nek}.}
  \bibinfo{year}{2018}\natexlab{}.
\newblock \showarticletitle{Bayesian online regression for adaptive direct
  illumination sampling}.
\newblock \bibinfo{journal}{\emph{ACM Transactions on Graphics (TOG)}}
  \bibinfo{volume}{37}, \bibinfo{number}{4} (\bibinfo{year}{2018}),
  \bibinfo{pages}{1--12}.
\newblock


\bibitem[Vorba et~al\mbox{.}(2014a)]%
        {SamplingVorba}
\bibfield{author}{\bibinfo{person}{Ji\v{r}\'{\i} Vorba},
  \bibinfo{person}{Ond\v{r}ej Karl\'{\i}k}, \bibinfo{person}{Martin \v{S}ik},
  \bibinfo{person}{Tobias Ritschel}, {and} \bibinfo{person}{Jaroslav
  K\v{r}iv\'{a}nek}.} \bibinfo{year}{2014}\natexlab{a}.
\newblock \showarticletitle{On-Line Learning of Parametric Mixture Models for
  Light Transport Simulation}.
\newblock \bibinfo{journal}{\emph{ACM Trans. Graph.}} \bibinfo{volume}{33},
  \bibinfo{number}{4}, Article \bibinfo{articleno}{101} (\bibinfo{date}{jul}
  \bibinfo{year}{2014}), \bibinfo{numpages}{11}~pages.
\newblock
\showISSN{0730-0301}
\urldef\tempurl%
\url{https://doi.org/10.1145/2601097.2601203}
\showDOI{\tempurl}


\bibitem[Vorba et~al\mbox{.}(2014b)]%
        {vorba2014line}
\bibfield{author}{\bibinfo{person}{Ji\v{r}\'{\i} Vorba},
  \bibinfo{person}{Ond\v{r}ej Karl\'{\i}k}, \bibinfo{person}{Martin \v{S}ik},
  \bibinfo{person}{Tobias Ritschel}, {and} \bibinfo{person}{Jaroslav
  K\v{r}iv\'{a}nek}.} \bibinfo{year}{2014}\natexlab{b}.
\newblock \showarticletitle{On-Line Learning of Parametric Mixture Models for
  Light Transport Simulation}.
\newblock \bibinfo{journal}{\emph{ACM Trans. Graph.}} \bibinfo{volume}{33},
  \bibinfo{number}{4}, Article \bibinfo{articleno}{101} (\bibinfo{date}{jul}
  \bibinfo{year}{2014}), \bibinfo{numpages}{11}~pages.
\newblock
\showISSN{0730-0301}
\urldef\tempurl%
\url{https://doi.org/10.1145/2601097.2601203}
\showDOI{\tempurl}


\bibitem[Walter et~al\mbox{.}(2005)]%
        {Lightcut2005}
\bibfield{author}{\bibinfo{person}{Bruce Walter}, \bibinfo{person}{Sebastian
  Fernandez}, \bibinfo{person}{Adam Arbree}, \bibinfo{person}{Kavita Bala},
  \bibinfo{person}{Michael Donikian}, {and} \bibinfo{person}{Donald~P.
  Greenberg}.} \bibinfo{year}{2005}\natexlab{}.
\newblock \showarticletitle{Lightcuts: A Scalable Approach to Illumination}.
\newblock \bibinfo{journal}{\emph{ACM Trans. Graph.}} \bibinfo{volume}{24},
  \bibinfo{number}{3} (\bibinfo{date}{jul} \bibinfo{year}{2005}),
  \bibinfo{pages}{1098–1107}.
\newblock
\showISSN{0730-0301}
\urldef\tempurl%
\url{https://doi.org/10.1145/1073204.1073318}
\showDOI{\tempurl}


\bibitem[Wang et~al\mbox{.}(2019)]%
        {wang2019fast}
\bibfield{author}{\bibinfo{person}{Yue Wang}, \bibinfo{person}{Soufiane Khiat},
  \bibinfo{person}{Paul~G. Kry}, {and} \bibinfo{person}{Derek Nowrouzezahrai}.}
  \bibinfo{year}{2019}\natexlab{}.
\newblock \showarticletitle{Fast Non-Uniform Radiance Probe Placement and
  Tracing}. In \bibinfo{booktitle}{\emph{Proceedings of the ACM SIGGRAPH
  Symposium on Interactive 3D Graphics and Games}} (Montreal, Quebec, Canada)
  \emph{(\bibinfo{series}{I3D '19})}. \bibinfo{publisher}{Association for
  Computing Machinery}, \bibinfo{address}{New York, NY, USA}, Article
  \bibinfo{articleno}{5}, \bibinfo{numpages}{9}~pages.
\newblock
\showISBNx{9781450363105}
\urldef\tempurl%
\url{https://doi.org/10.1145/3306131.3317024}
\showDOI{\tempurl}


\bibitem[Ward et~al\mbox{.}(1988)]%
        {RadianceOld}
\bibfield{author}{\bibinfo{person}{Gregory~J. Ward},
  \bibinfo{person}{Francis~M. Rubinstein}, {and} \bibinfo{person}{Robert~D.
  Clear}.} \bibinfo{year}{1988}\natexlab{}.
\newblock \showarticletitle{A Ray Tracing Solution for Diffuse
  Interreflection}.
\newblock \bibinfo{journal}{\emph{SIGGRAPH Comput. Graph.}}
  \bibinfo{volume}{22}, \bibinfo{number}{4} (\bibinfo{date}{jun}
  \bibinfo{year}{1988}), \bibinfo{pages}{85–92}.
\newblock
\showISSN{0097-8930}
\urldef\tempurl%
\url{https://doi.org/10.1145/378456.378490}
\showDOI{\tempurl}


\bibitem[Welling and Teh(2011)]%
        {welling2011bayesian}
\bibfield{author}{\bibinfo{person}{Max Welling} {and} \bibinfo{person}{Yee~Whye
  Teh}.} \bibinfo{year}{2011}\natexlab{}.
\newblock \showarticletitle{Bayesian Learning via Stochastic Gradient Langevin
  Dynamics}. In \bibinfo{booktitle}{\emph{Proceedings of the 28th International
  Conference on International Conference on Machine Learning}} (Bellevue,
  Washington, USA) \emph{(\bibinfo{series}{ICML'11})}.
  \bibinfo{publisher}{Omnipress}, \bibinfo{address}{Madison, WI, USA},
  \bibinfo{pages}{681–688}.
\newblock
\showISBNx{9781450306195}


\bibitem[Xiao et~al\mbox{.}(2020)]%
        {NeuralSS}
\bibfield{author}{\bibinfo{person}{Lei Xiao}, \bibinfo{person}{Salah Nouri},
  \bibinfo{person}{Matt Chapman}, \bibinfo{person}{Alexander Fix},
  \bibinfo{person}{Douglas Lanman}, {and} \bibinfo{person}{Anton Kaplanyan}.}
  \bibinfo{year}{2020}\natexlab{}.
\newblock \showarticletitle{Neural Supersampling for Real-Time Rendering}.
\newblock \bibinfo{journal}{\emph{ACM Trans. Graph.}} \bibinfo{volume}{39},
  \bibinfo{number}{4}, Article \bibinfo{articleno}{142} (\bibinfo{date}{July}
  \bibinfo{year}{2020}), \bibinfo{numpages}{12}~pages.
\newblock
\showISSN{0730-0301}
\urldef\tempurl%
\url{https://doi.org/10.1145/3386569.3392376}
\showDOI{\tempurl}


\end{thebibliography}

\appendix

\section{Metropolis-hastings}
Markov Chain Monte Carlo (MCMC) allows sampling from the posterior without computing the marginal. \citep{geyer1992practical}. Metropolis-Hastings (Metropolis), which we exploit in this work, is a specific implementation of MCMC \citep{chib1995understanding}. The Metropolis–Hastings algorithm can draw samples from any probability distribution with probability density $P(x)$, provided a function $h(x)$ proportional to the density $P(x)$. The Metropolis algorithm works by generating a sequence of sample values so that, as more samples are produced, the distribution of samples more closely approximates the desired distribution. These sample values are produced iteratively, meaning the next sample being dependent on the current sample (thus making the sequence of samples into a chain). Let $h(x)$ be a function that is proportional to the desired probability density function $P(x)$ (a.k.a. a target distribution). The Metropolis Markov Chain algorithm with random walk is defined as follows:

\begin{algorithm}[ht]
\DontPrintSemicolon
\caption{Random-walk algorithm}\label{alg:MetropolisStep}
\KwInput{$x^i$: Current state, $y^i$ : Probability of current state}
\KwInput{$\sigma$ : Step size or std-dev of Gaussian noise}
\KwOutput{$x^{i+1}$: Next state, $y^{i+1}$ : Probability of next state}
\SetKwFunction{MHS}{RandomWalk}
\SetKwProg{Fn}{function}{:}{}
  \Fn{\MHS{$x^i$, $y^i$}}
  {
        $x^{i + 1} \gets x^i + \mathcal{N}(\sigma)$ \tcp*{Propose a new state}
        $y^{i + 1} \gets h(x^{i + 1})$\;
        $\mu \gets \min\left\{\frac{y^{i + 1}}{y^i}, 1\right\}$ \tcp*{Compute acceptance ratio}
        $\epsilon \sim U(0,1)$ \tcp*{Sample uniform distribution}
        \If{$\epsilon > \mu$}
        {   \tcc{Reject proposed state}
            $x^{i + 1} \gets x^i$\;
            $y^{i + 1} \gets y^i$\;
        }
        \KwRet $x^{i+1}, y^{i+1}$\;
 }
\end{algorithm}

\textbf{Initialization:} Choose an arbitrary point $x^{i-1}$ as the initial observation in the sample-space and choose an arbitrary probability density $\mathcal{N}(x^{i}\mid x^{i-1})$ that suggests the next sample candidate $x^{i}$, given the previous sample value $x^{i-1}$. In our work, $\mathcal{N}$ is assumed to be symmetric. A usual choice is to let $\mathcal{N}(x^{i}\mid x^{i-1})$ be a Gaussian distribution centered at $x^{i-1}$, so that points closer to $x^{i-1}$ are more likely to be visited next, making the sequence of samples resemble a \textit{random walk} \citep{chib1995understanding}. The random walk algorithm is described in algorithm \ref{alg:MetropolisStep}.

\section{Probe compression \label{sec:probeCompression}}

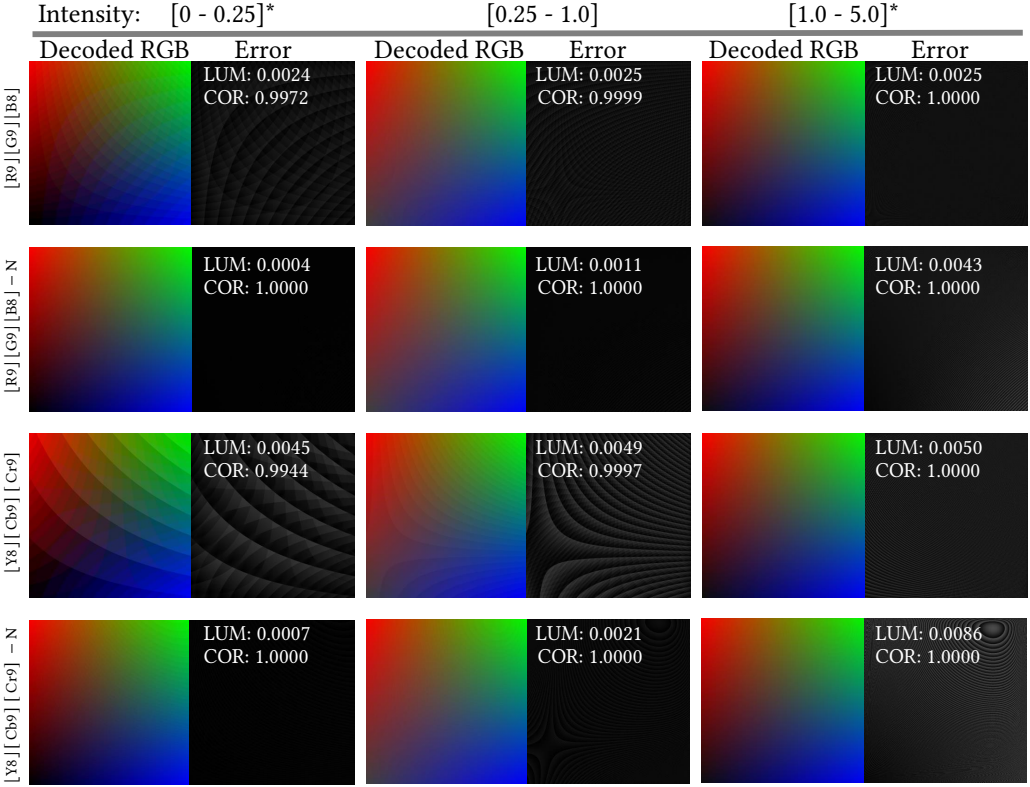
\begin{figure*}[t!]
\begin{tikzpicture}
    \node[anchor=south west,inner sep=0] at (0,0){\includegraphics[width=13.5cm, trim={0cm 20.0cm 0.1cm 0cm},clip]{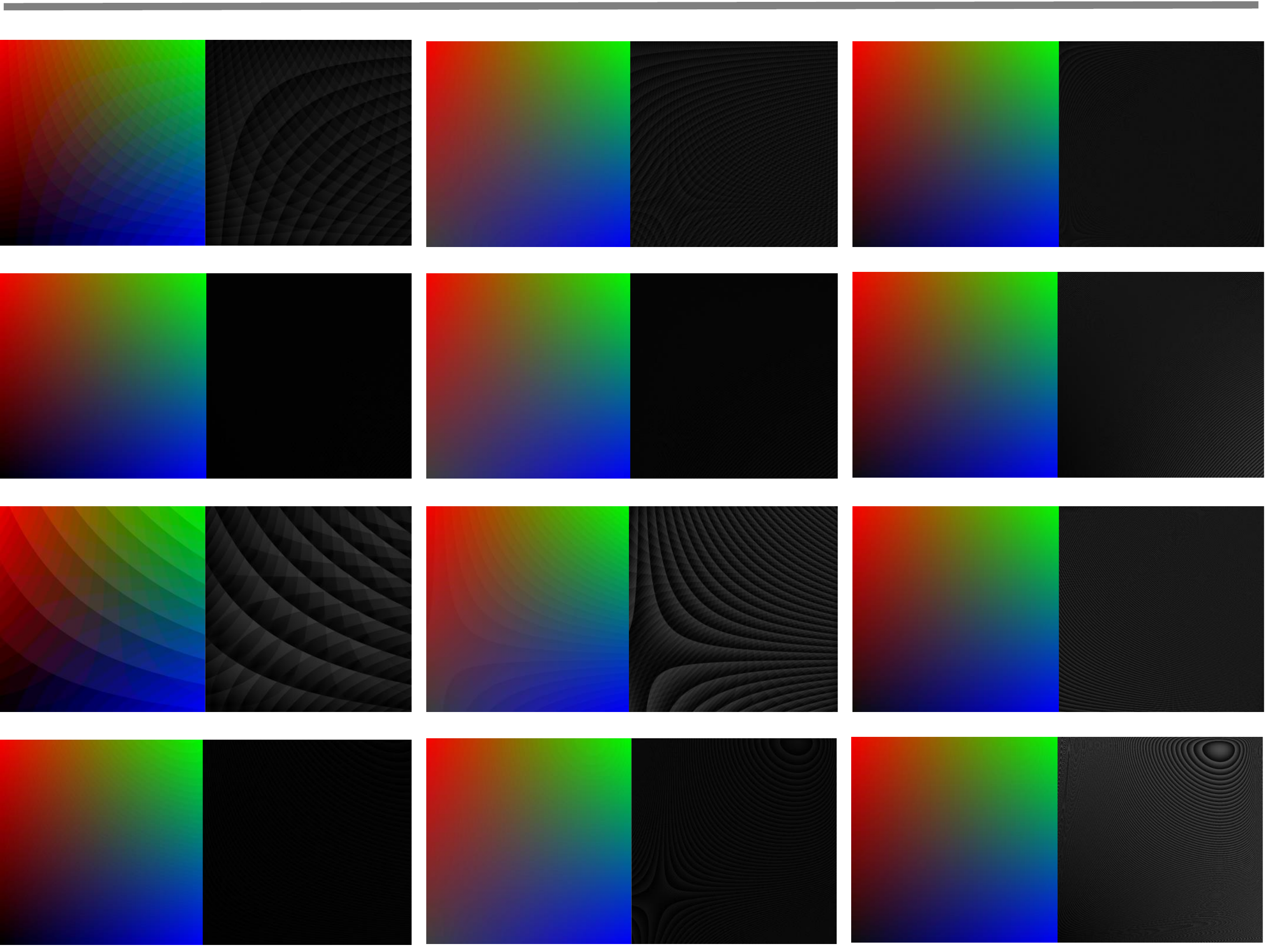}}; 
     \node[anchor=south west] at (0.0, 9.9) { \textcolor{black}{Intensity: \hspace{3mm}[0 - 0.25]*\hspace{28mm}[0.25 - 1.0]\hspace{25mm}[1.0 - 5.0]* } };
     
      \node[anchor=south west] at (0.0, 9.5) { \textcolor{black}{\hspace{0.2mm}Decoded RGB\hspace{6mm}Error\hspace{11mm}Decoded RGB\hspace{6mm}Error\hspace{11mm}Decoded RGB\hspace{7mm}Error } };
      
       \node[anchor=south west, rotate=90] at (0.0, -0.05) { \textcolor{black}{\tiny \hspace{0mm}$\lfloor \text{Y8} \rfloor [\,\text{Cb9}]\,[\,\text{Cr9}]\,-\text{N}$\hspace{7mm}$\lfloor \text{Y8} \rfloor [\,\text{Cb9}]\,[\,\text{Cr9}]\,$\hspace{7mm}$\lfloor \text{R9} \rfloor \lfloor \text{G9} \rfloor \lfloor \text{B8} \rfloor - \text{N}$\hspace{9mm}$\lfloor \text{R9} \rfloor \lfloor \text{G9} \rfloor \lfloor \text{B8} \rfloor$ } };
       
      \node[anchor=south west] at (2.2, 9.5-0.3) { \textcolor{white}{\footnotesize LUM: 0.0024\hspace{30mm}LUM: 0.0025\hspace{31mm}LUM: 0.0025}};
       \node[anchor=south west] at (2.2, 6.95-0.25) { \textcolor{white}{\footnotesize LUM: 0.0004\hspace{30mm}LUM: 0.0011\hspace{31mm}LUM: 0.0043}};
       \node[anchor=south west] at (2.2, 4.4-0.1) { \textcolor{white}{\footnotesize LUM: 0.0045\hspace{30mm}LUM: 0.0049\hspace{31mm}LUM: 0.0050}};
       \node[anchor=south west] at (2.2, 1.85) { \textcolor{white}{\footnotesize LUM: 0.0007\hspace{30mm}LUM: 0.0021\hspace{31mm}LUM: 0.0086}};
       
        \node[anchor=south west] at (2.2, 9.2-0.3) { \textcolor{white}{\footnotesize COR: 0.9972\hspace{30.5mm}COR: 0.9999\hspace{31mm}COR: 1.0000}};
       \node[anchor=south west] at (2.2, 6.65-0.25) { \textcolor{white}{\footnotesize COR: 1.0000\hspace{30.5mm}COR: 1.0000\hspace{31mm}COR: 1.0000}};
       \node[anchor=south west] at (2.2, 4.1-0.1) { \textcolor{white}{\footnotesize COR: 0.9944\hspace{30.5mm}COR: 0.9997\hspace{31mm}COR: 1.0000}};
       \node[anchor=south west] at (2.2, 1.55) { \textcolor{white}{\footnotesize COR: 1.0000\hspace{30.5mm}COR: 1.0000\hspace{31mm}COR: 1.0000}};
\end{tikzpicture}
\vspace{-20pt}
\caption{Figure comparing 26-bit color encodings on slices of the 3D color-space with dynamic range. We compare the reconstruction error measured in Luminance and Color Correlation with RGB32f reference. The log-non-linear encodings marked with - N suffix shifts the bit error from lower to higher intensities - which are less frequent in indirect illumination. $\lfloor\rfloor$ and $[\,]\,$ denotes round-low and round-nearest quantizations respectively. * Color map visualizations are normalized.}
\label{fig:rgbEnc}
\end{figure*}

We tested several 26-bit encoding and settled on a non-linear RGB encoding represented by $\lfloor \text{R9} \rfloor \lfloor \text{G9} \rfloor \lfloor \text{B8} \rfloor - \text{N}$ in figure \ref{fig:rgbEnc}. In this encoding, the RGB color is first passed through a logarithmic non-linearity as per equation \ref{eq:rgbEnc} such that the quantization errors are distributed evenly across intensities. We perform a \textit{round-to-lowest-integer} ($\lfloor\rfloor$) quantization for all channels, although \textit{round-to-nearest-integer} ($[\,]\,$) is more accurate.  Our quantization scheme ensures the moving-average updates produce dark colors when the intensity of new samples are low. In a \textit{round-to-nearest} setting, due to a \textit{round-up} error, the colors may never go to zero. Interestingly, YCbCr encoding allows \textit{round-to-lowest} for the Y channel and round \textit{round-to-nearest} for Cb and Cr channels, however, they perform poorly in both luminance and color preservation metrics as shown in figure  \ref{fig:rgbEnc}.

The parameters in equation \ref{eq:rgbEnc} are obtained by performing a grid search minimizing the reconstruction error w.r.t RGB32f reference across various color and intensity combinations as shown in figure \ref{fig:rgbEnc}. Luminance error is the r.m.s. value of the difference between the two color-maps. Color accuracy is measured using a normalized dot product between the two flattened color-maps.

\end{document}